\documentclass[review]{elsarticle}

\usepackage{lineno,hyperref}
\modulolinenumbers[5]

\journal{Journal of \LaTeX\ Templates}

\usepackage{amsmath}

\begin{document}

\begin{frontmatter}

\title{Heterogeneous wealth distribution, round-trip trading and the emergence of volatility clustering in Speculation Game}


\author[mymainaddress,mysubaddress]{Kei Katahira\corref{mycorrespondingauthor}}
\cortext[mycorrespondingauthor]{Corresponding author.}
\ead{k.katahira@scslab.k.u-tokyo.ac.jp}

\author[mymainaddress]{Yu Chen}

\address[mymainaddress]{Graduate School of Frontier Sciences, The University of Tokyo, 5-1-5 Kashiwanoha, Kashiwa-shi, Chiba-ken 277-8563, Japan}
\address[mysubaddress]{Research Fellow of Japan Society for the Promotion of Science}

\begin{abstract}
This study is a detailed analysis of Speculation Game, a minimal agent-based model of financial markets, in which the round-trip trading and the dynamic wealth evolution with variable trading volumes are implemented. Instead of herding behavior, we find that the emergence of volatility clustering can be induced by the heterogeneous wealth distribution among traders. In particular, the spontaneous redistribution of market wealth through repetitions of round-trip trades can widen the wealth disparity and establish the Pareto distribution of the capital size. In the meantime, large fluctuations in price return are brought on by the intermittent placements of the relatively big orders from rich traders. Empirical data are used to support the scenario derived from the model. 
\end{abstract}


\end{frontmatter}

\section{Introduction}
The time series of financial asset returns are known to have a set of nontrivial quantitative and qualitative properties collectively called as the stylized facts \cite{cont2001empirical}. One of the representative features is {\it volatility clustering}, a jargon related to the inclination of large fluctuations in price returns to form clusters, which results in a long time tail (often observed as either a power-law or a logarithmic function) in the decay of autocorrelation of volatility \cite{mantegna1999introduction, rydberg2000realistic, zumbach2007riskmetrics}. Moreover, those clusters burst intermittently over a wide spectrum of time scales \cite{cont2001empirical, mantegna1999introduction, lux1999scaling} so that the temporal structure of price returns can be considered as fractals featuring with the absence of characteristic time scale \cite{mandelbrot1982fractal}. As the volatility clustering can be observed in the different markets and instruments, it attracts the interest from many researchers, especially econophysicists, who are enthusiastic about discovering such a universality class and comprehending the mechanism behind.

Stochastic processes, which are standard models in the financial and econometric fields, have been used to describe the volatility clustering at the aggregated level. The representative models are the autoregressive conditional heteroscedasticity (ARCH) and generalized autoregressive conditional heteroscedasticity (GARCH) processes, both of which can reproduce the aperiodic bursts in returns \footnote{Nevertheless, the autocorrelation of volatility of times series data yielded by GARCH decays exponentially \cite{zumbach2007riskmetrics}.} \cite{mantegna1999introduction, bollerslev1986generalized}. Typically, these models are useful for practical analyses of empirical data, whereas they do not take account of endogenous factors at the microscopic level of financial markets.

On the other hand, bottom-up approaches, such as agent-based models, are favored in the econophysics field because the cause for the emergence of volatility clustering can be captured more easily with the explicit description of the interaction among traders. Several types of very simple agent-based models (so-called toy models) were constructed for the detailed analysis of financial markets. The representative ones can be listed as the percolation model \cite{cont2000herd, stauffer1998crossover}, Ising (spin) model \cite{sornette2006importance, zhou2007self, kaizoji2002dynamics, sornette2014physics}, Sznajd model \cite{sznajd2002simple}, the grand canonical Minority Game (GCMG) \cite{challet2001stylized, challet2001games, bouchaud2001universal, challet2003criticality}, and so on. The findings from the studies with these models revealed that the phenomenon volatility clustering may be originated in the synchronization of traders, namely the {\it herding behavior} \cite{manuca2000structure}, which is the human tendency of following the actions of others \footnote{GCMG, however, the reproduction of volatility clustering highly depends on the random initialization of the strategy tables \cite{slanina2013essentials}. The autocorrelation of volatility shown in \cite{challet2001stylized} is indeed the failed case. Even when the volatility clustering turns up in this model, the tail length of its positive autocorrelation may vary quite a lot (ranging from shorter than 50-time lags to longer than 500-time lags).}. 

Meanwhile, there remains a possibility that some other mechanisms can also lead to the emergence of volatility clustering, considering that those toy models might overlook other fundamental microstructures of the market. Furthermore, although the inference of the contribution of herding behavior toward the irregular price bursts seems reasonable, it is hard to measure, even in a qualitative sense, the degree and the extent of herding in the real financial markets. Hence, it is natural and absolutely necessary to think about other potential mechanisms working for the emergence of volatility clustering. In fact, Maslov model \cite{maslov2000simple} has revealed that the continuous double auction price formation system alone can induce the presence of fat-tailed distribution and volatility clustering in price fluctuations. 

As a latent cause for the emergence of volatility clustering, we particularly pay attention to the combination of round-trip trading and dynamic wealth whose effects has not been clarified yet. While none of the introduced simple models concerns these details seriously, there are some prior toy models equipped with either one of the elements. For example, the concept of round-trip transactions has been installed in the \$-game \cite{andersen2003game} to evaluate trading strategies, but they are not implemented in agents' trading actions \footnote {Although agent's wealth in the \$-game changes dynamically, her/his order size is kept constant as one.}. Ferreira and Marsili \cite{ferreira2005real} introduced a round-trip trade mechanism into Minority Game, Majority Game and \$-game under the restriction of a successive two-step transaction. In the pattern game \cite{challet2008inter}, round-trip payoffs are incorporated by taking the reaction time of order placement into account. As for the dynamical wealth with variable trading volumes, there are extended GCMG models in which the wealth diversity and the varying investment sizes are considered \cite{jefferies2001market}. Minority Game with dynamical capital (MGDC) \cite{challet2001minority, galla2009minority} allows the fractional investment of agents’ total capital. Note that dynamic capital is significant to bring on heavy tails of price return, but it can hardly reproduce the distinctive volatility clustering in MGDC without the inductive learning (see more details in Subsubsection \ref{learning}). Lastly, it is worth mentioning that both the evolving money and the variable order size are implemented in the Patzelt-Pawelzik model \cite{patzelt2013inherent}.

Based on these previous studies, we developed {\it Speculation Game} \cite{katahira2019development}, which is an adaptive agent model allowing both an explicit description of round-trip trading and a dynamic evolution of traders' wealth. In Speculation Game, round-trip trading is defined as a trading process in which an agent, responding to a set of signals received from the price movements, opens a position with a buy/sell order first, and closes the position with the reverse order later. At other times, agents either hold a position in the ongoing round-trip trade or idle without any positions in the market. The sophistication of model building renders Speculation Game prominent  in reproducing 10 of 11 the well-known stylized facts for financial time series \cite{cont2001empirical} including the volatility clustering and the heavy tails, which are most concerned in the development of a model for the financial markets \footnote {See \cite{katahira2019development} for the reproducibility of Speculation Game for other stylized facts.}. Nevertheless, the detailed scenario of the emergence of stylized facts in Speculation Game is not clarified yet because the enhancement of reality also makes the analysis of model nontrivial. 

In this paper, the detailed mechanisms for Speculation Game working on the emergence of the volatility clustering is analyzed through a series of simulation study. In particular, the built-up of heterogeneous distribution of traders' wealth through the round-trip trades, together with the realization of variable investment size, are revealed as another possible scenario for the emergence of financial stylized facts. 

The remainder of this paper is organized as follows. The next section describes the main ingredient of Speculation Game, while the one after the next presents the emerging mechanism of volatility clustering. Finally, the last section concludes the study.

\section{Speculation Game}
\subsection{Model building}
Speculation Game is a repeated game in which players compete with each other to increase wealth by capital gains through round-trip trades. The model was constructed by applying the structure of Minority Game \cite{challet1997emergence, challet2005minority} including history, memory, and strategy table for decision making of the players. Moreover, two more distinctive parameters, namely board lot amount and cognitive threshold, are introduced to enable the players to place orders with variable trading volumes and to allow the history to carry both information of the direction and the magnitude of price change. Another unique point is that the game proceeds with alternations between realistic and cognitive worlds \footnote{See Fig. 1 of \cite{katahira2019development} for further understanding of the framework.}, which represents the human tendency in handling complex information through simplification.

In Speculation Game, $N$ players participate in a gamified market, with initial market wealth and $S$ strategy tables. At discrete time $t$, player $i$ uses her best strategy $j^*$ ($\in S$) to take an action $a_i^{j^*}(t)$ from three options, that is buy ($=1$), sell ($=-1$), or hold (idle) ($=0$). When the player submits a buy or sell (short selling is allowed) order following strategy $j^*$, the quantity of order $q_i(t)$ is decided with her market wealth $w_i(t)$ and the board lot amount $B$, the latter of which describes the ease in placing orders with multiple quantities, 
\begin{equation}
q_i(t) = \lfloor \frac{w_i(t)}{B} \rfloor,
\label{eq1}
\end{equation}
where $\lfloor\cdots\rfloor$ stands for the flooring operator. Note that the closing quantity $q_i(t)$ is required to be the same as the opening one $q_i(t_0)$ in a round-trip trade. Also, the player's initial market wealth $w_i(0)$ is decided with a uniformly distributed random number $U[0,100)$ as below to enable her to order one unit at least:
\begin{equation}
w_i(0) = \lfloor B+U[0,100) \rfloor.
\label{eq2}
\end{equation}
If the wealth of a player decreases an amount to $w_i(t)<B$ as a result of round-trip trade, the player will be forced to leave the market and substituted by a new player, whose market  wealth is similarly decided according to Eq. \ref{eq2}.

Letting the initial asset price to be $p(0)=100$, the market price change $\Delta p$ is the order imbalance equation proposed in \cite{cont2000herd}: 
\begin{equation}
\Delta p = p(t) - p(t-1) = \frac{1}{N}\sum_{i=1}^{N}a_i^{j^*}(t)q_i(t).
\label{eq3}
\end{equation}
The quantized price movement $h(t)$ is decided by the magnitude correlation between the price change $\Delta p$ and the cognitive threshold $C$, the latter of which is a threshold value used by the players to recognize a big price move: 
\begin{equation}
h(t) = \begin{cases}
2\ ({\rm largely\ up}) & {\rm if}\ \Delta p>C,\\
1\ ({\rm up}) & {\rm if}\ C \geq \Delta p>0,\\
0\ ({\rm stay}) & {\rm if}\ \Delta p=0,\\
-1\ ({\rm down}) & {\rm if}\ -C \leq \Delta p<0,\\ 
-2\ ({\rm largely\ down}) & {\rm if}\ \Delta p<-C.
\end{cases}
\label{eq4}
\end{equation}
Hence, the history $H(t)$ is a quinary time series in the players' cognitive world recording the past price movements. To take an action according to the best strategy $a_i^{j^*}(t)$, the player with memory $M$ at first take the reference of the last $M$ digits of history $H(t)$. Next, she looks up strategy $j^*$ to obtain a recommended action corresponding to the historical pattern. However, when the recommendation has no change from the opening order $a_i^{j^*}(t_0)$, the player will hold the opened position. With such trading rules, the market signal based position open/close is enabled for the player to accomplish a round-trip trade with a single position as well as the variable holding and idling periods. Note that the constraint for the completion of round-trip trades also impedes the spontaneous synchronized actions among the players (i.e., the herding behavior) \footnote{See Appendix A in \cite{katahira2019development} for further details.}.

To determine the best strategy for the reference of next action, all the strategies are evaluated through virtual round-trip trades in the background similarly to the way applied to the strategy in use. Performances of the strategies are assessed in terms of the accumulated strategy gains $G_i^j(t)$ ($j \in S$) calculated with the cognitive price $P(t)$ corresponding to the quantized information. Letting $P(0) = 0$, $P(t)$ is updated as 
\begin{equation}
P(t) = P(t-1)+h(t).
\label{eq5}
\end{equation}
The gain of strategy $j$ in a round-trip trade $\Delta G_i^j(t)$ can read as
\begin{equation}
\Delta G_i^j(t) = a_i^j(t_0)(P(t) - P(t_0)),
\label{eq6}
\end{equation}
and the accumulated strategy gain $G_i^j(t)$ is measured by:
\begin{equation}
G_i^j(t) = G_i^j(t_0) + \Delta G_i^j(t).
\label{eq7}
\end{equation}
Whenever the accumulated gain of the strategy in use $G_i^{j^*}(t)$ is updated, all the accumulated strategy gains $G_i^j(t)$ will be reviewed to renew strategy $j^*$ with the best performed one. If the renewed best strategy happens to be one of the unused strategies with which a virtual trade is ongoing, the virtual position will be closed immediately (i.e., the virtual round-trip trade is aborted forthwith) before the player switches to this new best strategy at the next time step. Note that the evaluating system is developed by considering that the investing strategies should be evaluated with capital gains and losses by round-trip trades, as Katahira and Akiyama pointed out \cite{katahira2017}.

Since the self-financing assumption is not made in Speculation Game, when a round-trip trade is closed, the player's market wealth $w_i(t)$ is updated with an investment adjustment $\Delta  w_i(t)$, which is the conversion of strategy gain into market gain by taking the trading volume $q_i(t)$ into consideration,
\begin{equation}
w_i(t) = w_i(t_0) + \Delta  w_i(t) = w_i(t_0) + f(\Delta G_i^{j^*}(t)q_i(t_0)).
\label{eq8}
\end{equation}
Here $f$ can be an arbitrary function. In this study, $\Delta  w_i(t) = \Delta G_i^{j^*}(t)q_i(t_0)$ is used for the simplicity. 

\subsection{Phase diagram}
\begin{figure}[tbhp]
\begin{center}
\includegraphics[width=.6\textwidth]{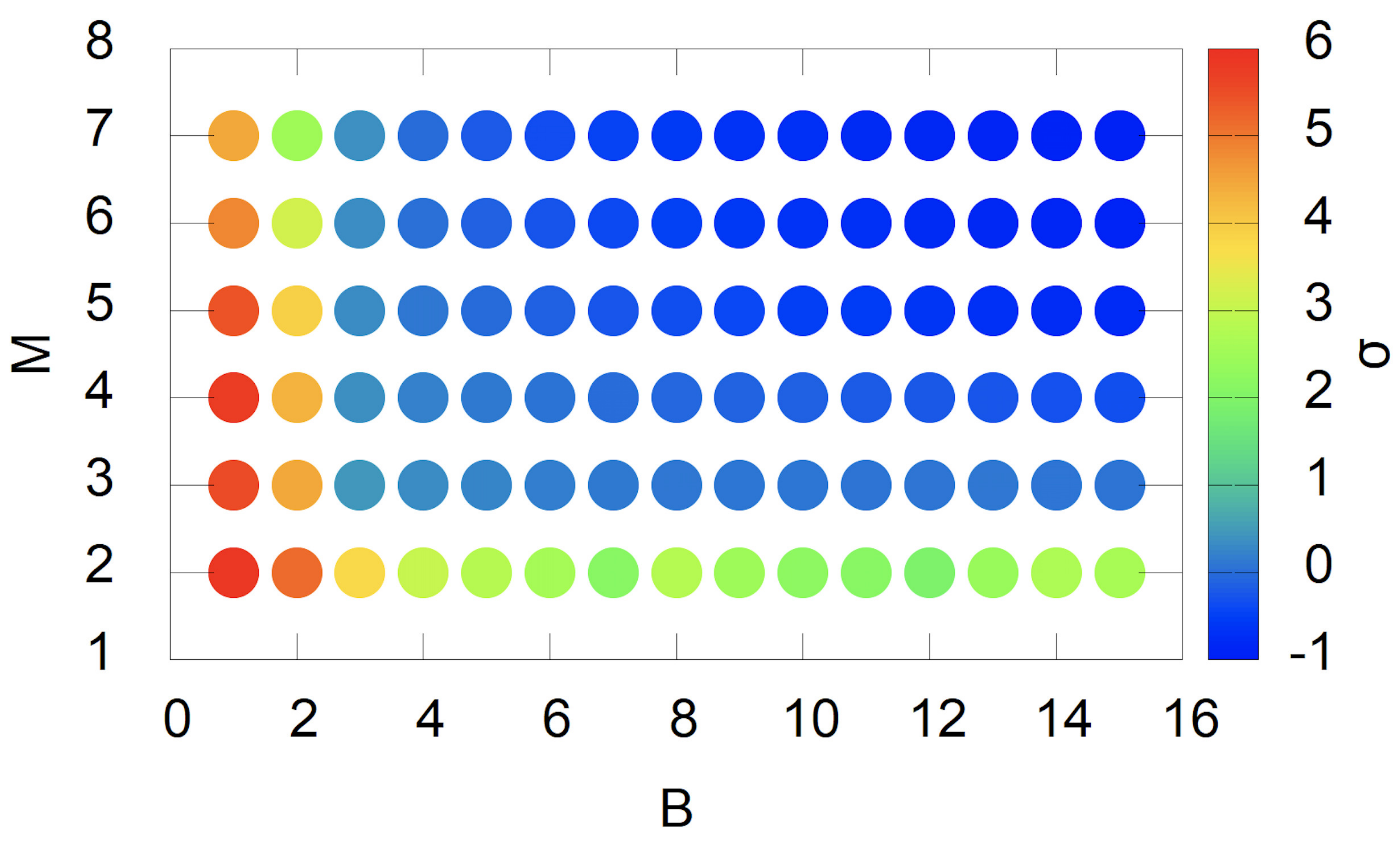}
\end{center}
\caption{The $M$-$B$ phase diagram of averaged $\sigma$ in log scale ($N=1,000$, $S=2$, $C=3$). The color-bar numbers are exponents of a logarithm. The diagram is plotted from 100-trial simulations with 50,000 iterations for each pair of parameters.}
\label{fig1}
\end{figure}

Under specific parameter settings, Speculation Game will get into the extreme state in which $\Delta p$ bursts irregularly with tremendous amplitudes \footnote{See Fig. B.1 in \cite{katahira2019development}.}. The extreme state can be differentiated from the stable state by inspecting the $M$ (memory) - $B$ (board lot amount) phase diagram of $\sigma$, where $\sigma$ is the standard deviation of market price change. Fig. \ref{fig1} shows such a diagram, where the color of circular dots represents 100-trial averaged $\sigma$ in log scale. As either $M$ or $B$ decreases, the standard deviation of $\Delta p$ increases monotonically. Especially when $M$ or $B$ is critically reduced from three to two, standard deviation $\sigma$ grows drastically (more than two decades). Accordingly, dots in (pale $\sim$ vivid) blue ($M\geq3　\cap B\geq3$) can be defined as the stable states, while the rest of dots as the extreme states ($\sigma > 10$).

Since the volatility clustering emerges only in the stable states, we start the analysis with the cases of $M\geq3$, particularly the comparison between the cases of $M=5$ and $M=7$, in the next subsubsection. Note that the simulation results along different $M$ shown in the following sections are obtained with the rest parameters fixed as $N=1,000$, $S=2$, $B=9$, $C=3$. Discussion on the effects of different values of these parameters can be found in the Appendices.

\section{Emergence of volatility clustering}
\subsection{The role of wealth distribution}
\subsubsection{Placement of big orders}
\begin{figure}[tbhp]
\begin{center}
(a)\includegraphics[width=0.85\textwidth]{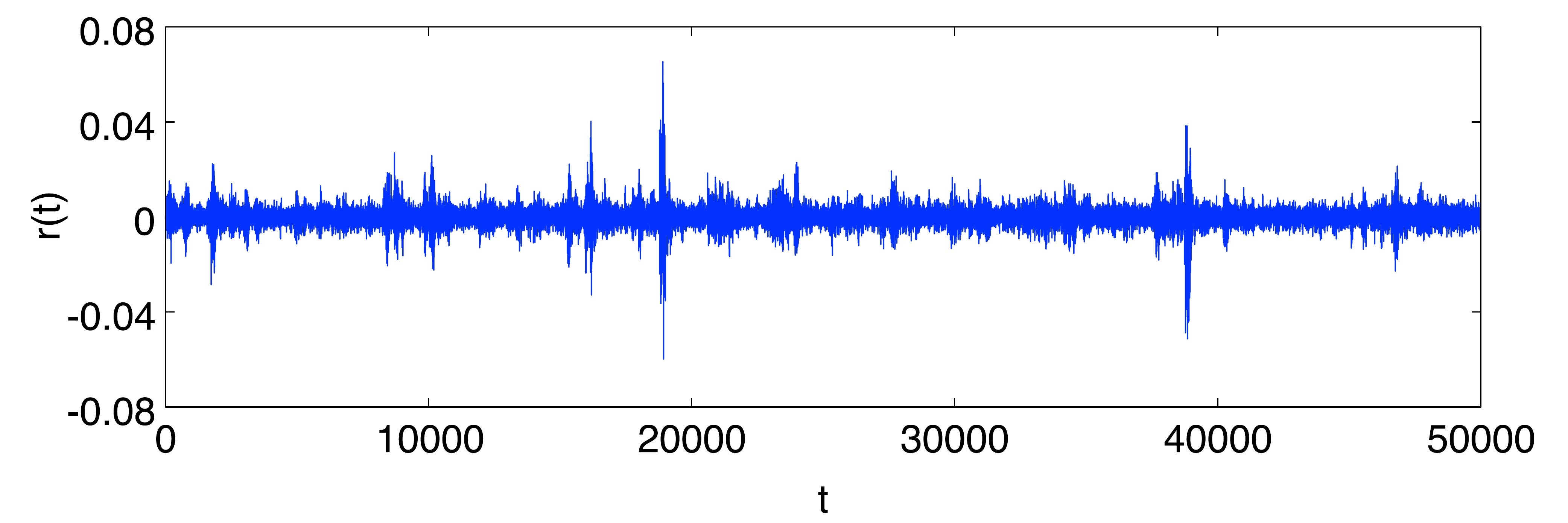}
(b)\includegraphics[width=0.85\textwidth]{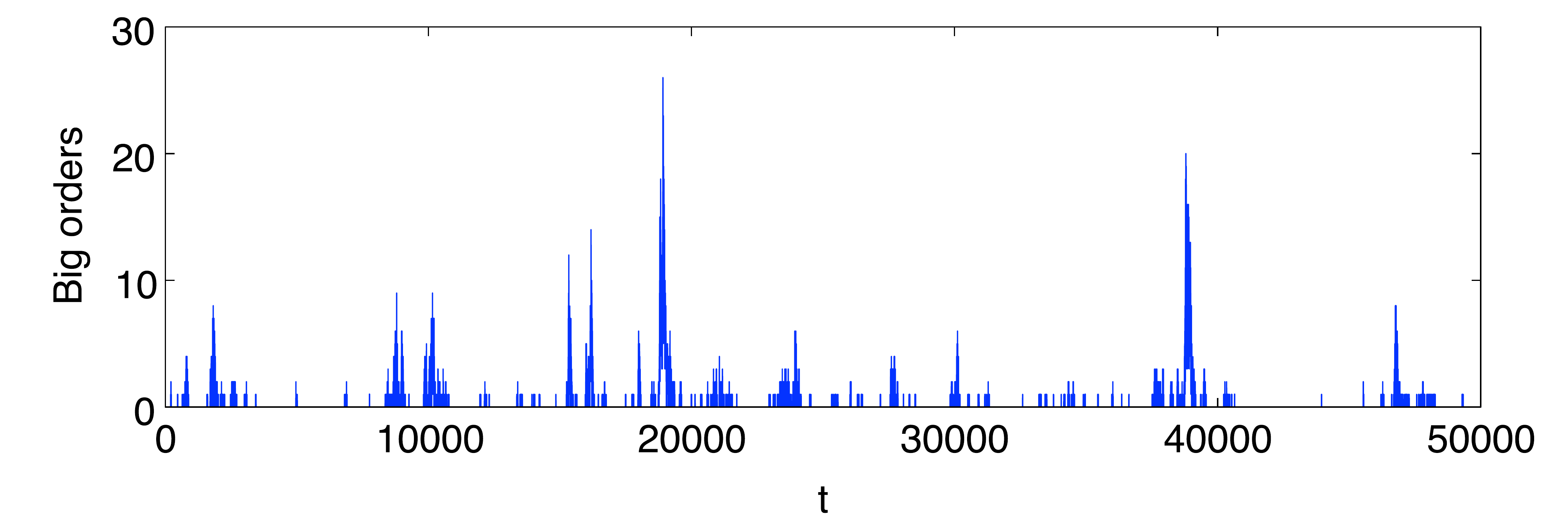}
(c)\includegraphics[width=0.85\textwidth]{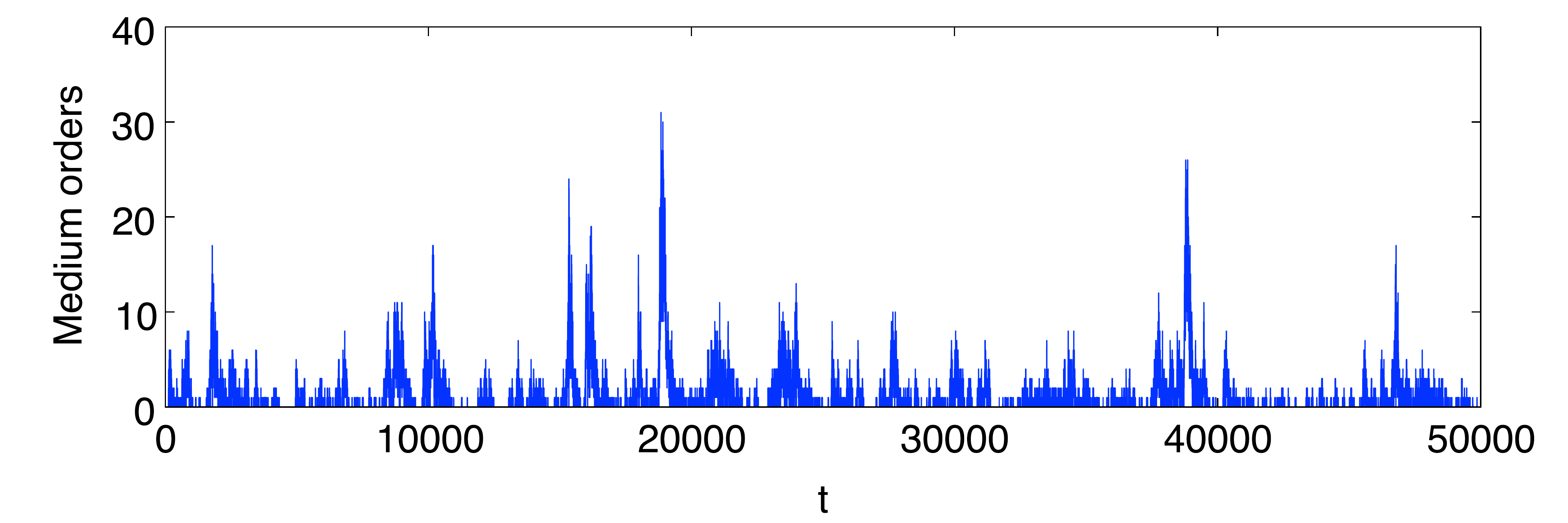}
(d)\includegraphics[width=0.85\textwidth]{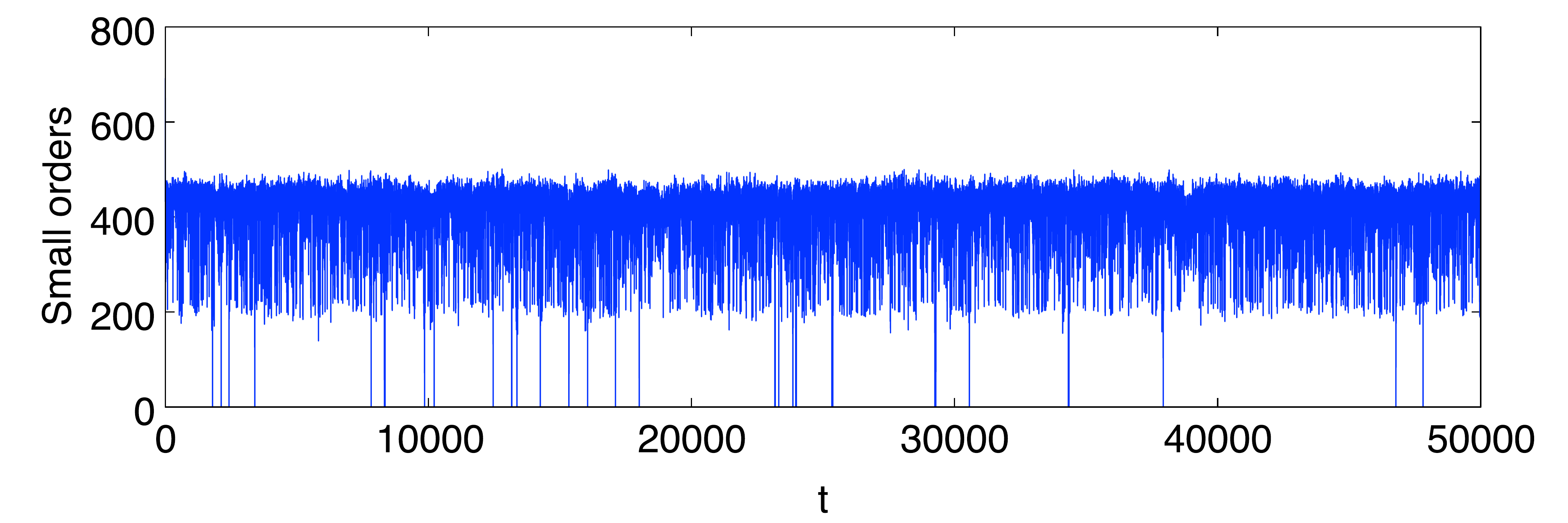}
\end{center}
\caption{Time series of (a) market returns and of the number of (b) big orders, (c) medium orders, and (d) small orders in the case of $M=5$.}
\label{fig2}
\end{figure}

\begin{figure}[tbhp]
\begin{center}
(a)\includegraphics[width=0.85\textwidth]{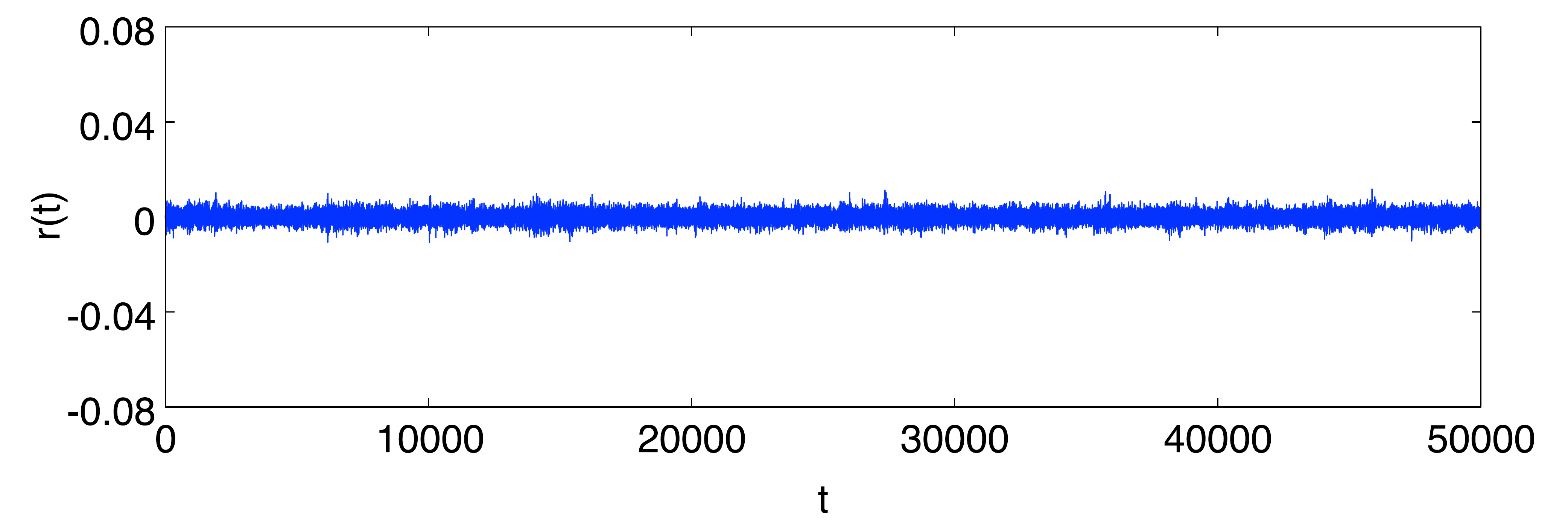}
(b)\includegraphics[width=0.85\textwidth]{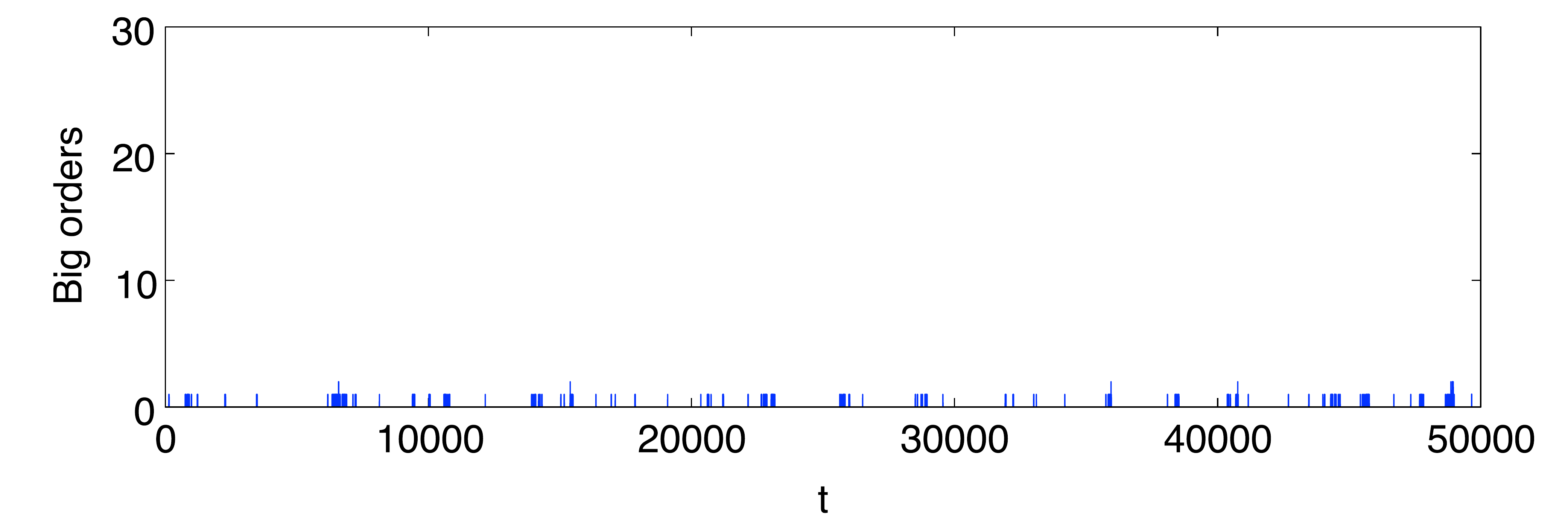}
(c)\includegraphics[width=0.85\textwidth]{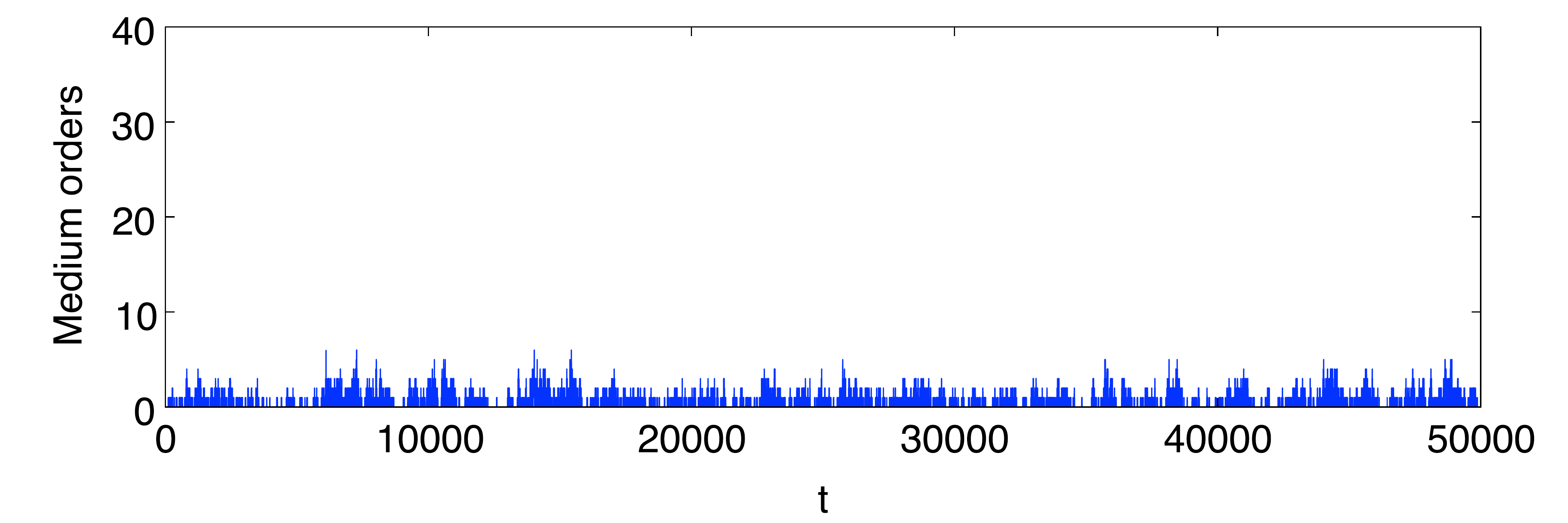}
(d)\includegraphics[width=0.85\textwidth]{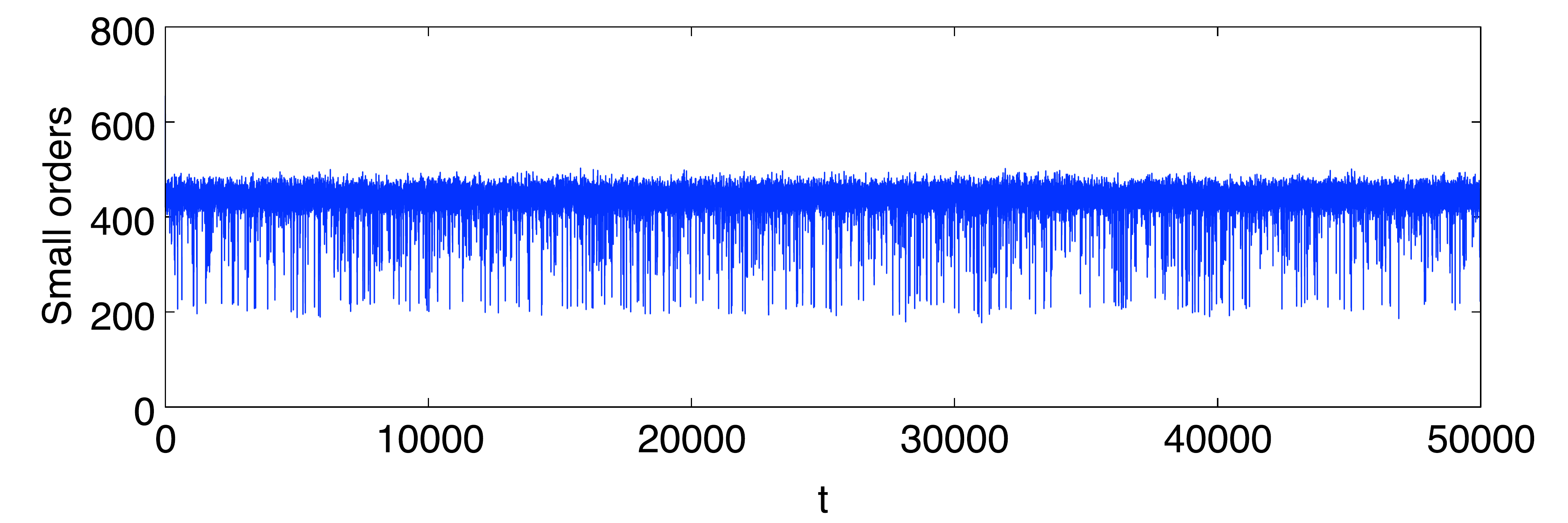}
\end{center}
\caption{Time series of (a) market returns and of the number of (b) big orders, (c) medium orders, and (d) small orders in the case of $M=7$.}
\label{fig3}
\end{figure}

The appearance of time series $r(t) (= {\rm ln}(p(t)) - {\rm ln}(p(t-1)))$ looks quite different depending on the value of $M$, in particular, the volatility clusters present more manifestly with smaller $M$. In the case of $M=5$, the intermittent burst of time series $r(t)$ is displayed in panel (a) of Fig. \ref{fig2}. In the case of $M=7$, large fluctuations in $r(t)$ are absent, see panel (a) of Fig. \ref{fig3}. This distinctive characteristics comes from the size of orders. Time series of the number of orders in different sizes are shown in panels (b)$\sim$(d) of Fig. \ref{fig2} and Fig. \ref{fig3} correspondingly. Big orders plotted in panel (b) are classified as $q_i(t)>100$, medium orders in panel (c) as $50<q_i(t)\leq100$, and small orders in panel (d) as $q_i(t)\leq50$. As Fig. \ref{fig2} shows, in the case of $M=5$, big and medium orders turn up at times when $r(t)$ fluctuates largely, while the number of small orders is almost constant just below $N/2$. In contrast, in the case of $M=7$, as it can be confirmed in Fig. \ref{fig3}, no distinguishable intermittent augmentations of big and medium orders can be found, whereas there is no big change regarding the incessant presence of small orders. The only exception is the existence of the sparse vacant points without any order placements in the case of $M=5$. 

These results indicate that the relatively big and medium order placements play a key role in the emergence of volatility clustering. According to Eq. \ref{eq1}, the quantity of order $q_i(t)$ is decided proportionally by the player's market wealth $w_i(t)$, which means that a wealthier player would place an order with a bigger trading volume. In other words, the intermittent time series of price returns implies the ample existence of wealthier players when $M$ gets smaller.

\subsubsection{Appearance of rich players}
\begin{figure}[tbhp]
\begin{center}
\includegraphics[width=.6\textwidth]{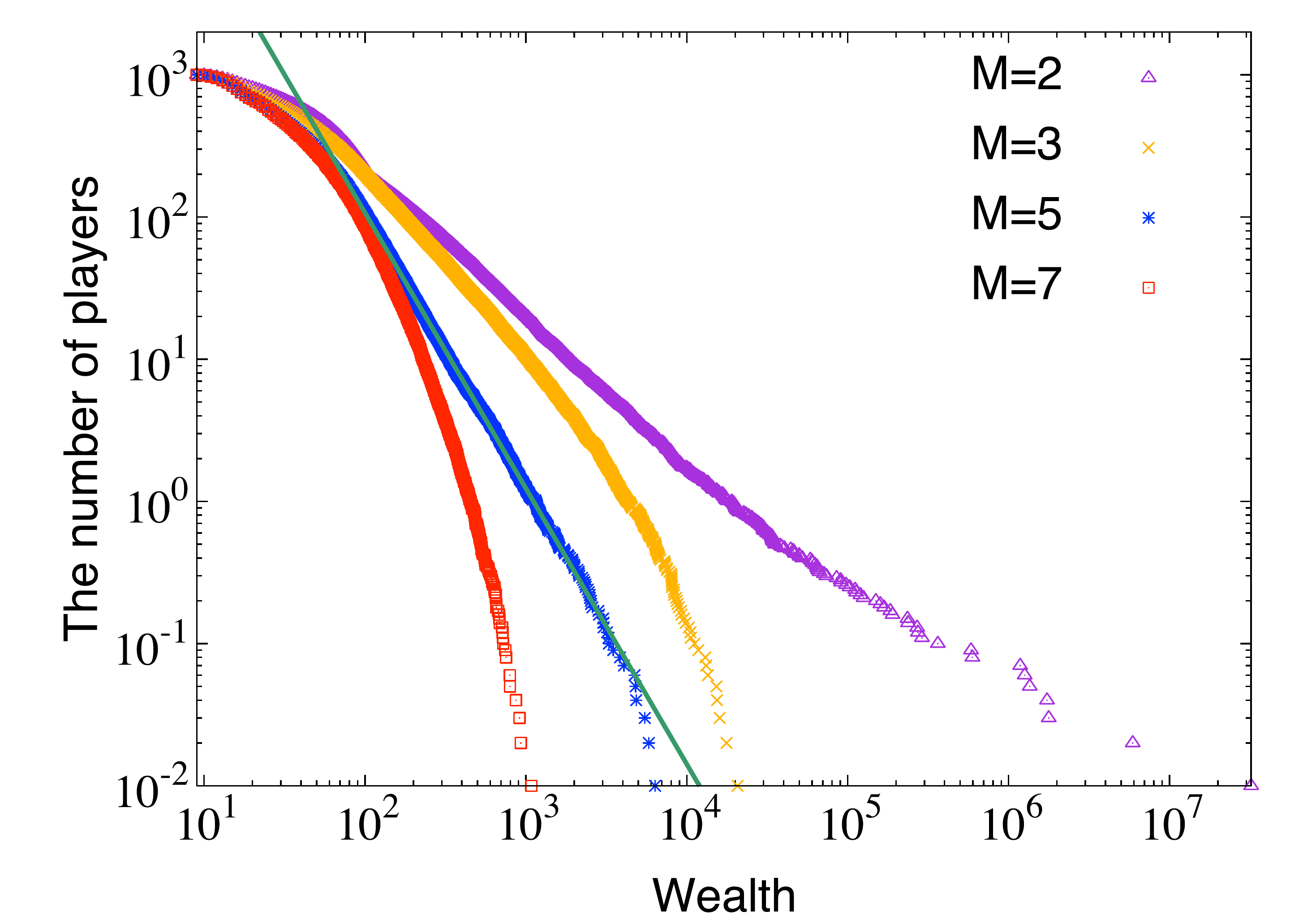}
\end{center}
\caption{A snapshot of averaged wealth distributions at $t=50,000$. These distributions are obtained from 100 trials of the simulation. States of system may be described more distinctively by the tail conditions of players' wealth such as stable state ($M=7$: a deflated tail), critical state ($M=5$: a power-law tail), super-critical state ($M=3$: an inflated tail), and extreme state ($M=2$: a super-inflated tail).}
\label{fig4}
\end{figure}

This implication of activities of rich players can be verified from a different aspect if one takes a snapshot of wealth distribution among the players after the elapse of a transition period for the system. Fig. \ref{fig4} is the log-log plot of averaged complementary cumulative distributions of wealth at $t=50,000$ in cases of $M=5$ (blue asterisk dots) and $M=7$ (red square dots), and the cases of $M=2$ (purple triangle dots) and $M=3$ (orange cross dots) for a further comparison \footnote {See Appendix A for the time series plots of returns and order placements in different sizes in the cases of $M=2$ and $M=3$.}. As Fig. \ref{fig4} shows, the blue colored distribution has a fatter tail than the red one has, meaning a larger number of rich players exists in the case of $M=5$ than that in the case of $M=7$. In other words, repetitions of the round-trip trades performed with a smaller $M$ would spontaneously generate a wider distribution of wealth among players \footnote {Smaller $B$ could also widen the wealth disparity among players. See Appendix B.}. This observation also applies to the cases of $M=2$ and $M=3$ as the fatter tails of purple and orange dots show. Based on results in Figs. \ref{fig2} - \ref{fig4}, one can say that the growth of volatility is brought by those wealthier players placing orders with big and medium quantities. 

Note that the complementary cumulative distribution of wealth in the case of $M=5$ follows Pareto's law \cite{persky1992retrospectives} with a tail index $\alpha=1.94$, which is indicated by the green line in Fig. \ref{fig4}. The lower cut-off of asymptotic power-law function is decided as $x_{min}=137$ by the ``poweRlaw'' package \cite{Gillespie2015}, developed by Clauset, Shalizi, and Newman \cite{clauset2009power}. 

\subsubsection{Empirical evidence}
\begin{figure}[tbhp]
\begin{center}
\includegraphics[width=.6\textwidth]{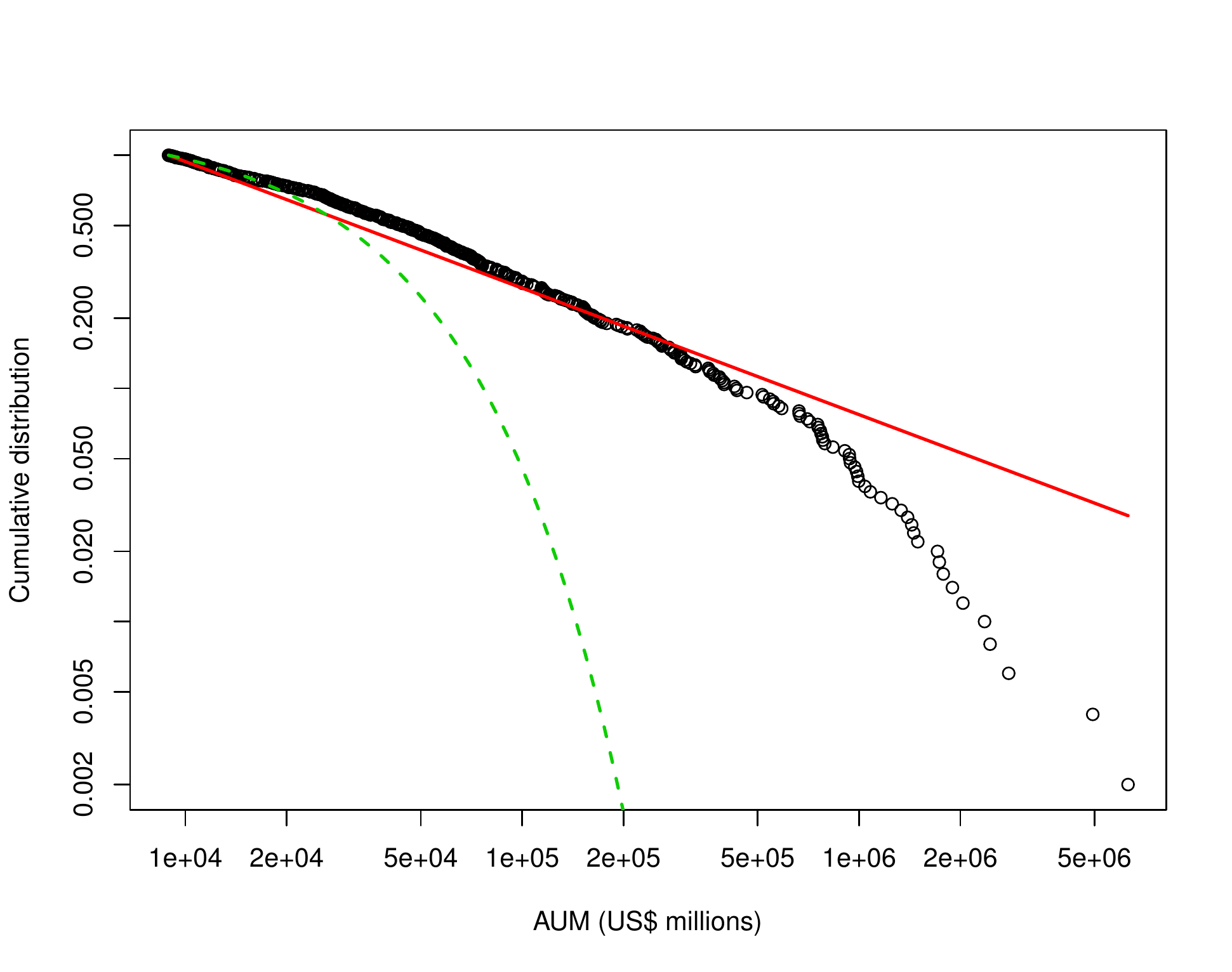}
\end{center}
\caption{Complementary cumulative distributions of AUM of the world's  500 largest asset managers in 2017. The scale is in US\$ millions.}
\label{fig5}
\end{figure}

Pareto's distribution of wealth even holds in the actual financial market as shown in Fig. \ref{fig5}, which is the log-log plot of complementary cumulative distributions of total discretionary assets under management (AUM) of the world's 500 largest asset managers in 2017. The AUM data is taken from the report done by Willis Towers Watson \cite{watson2018world}. The lower cut-off of this asymptotic power-law function ($x_{min}=8,914$) is also found with the ``poweRlaw'' package \cite{Gillespie2015}. The red line is a fitted power-law function with $\alpha = 1.54$ ($\alpha = 0.689$ when the vertical axis is the number of asset managers), whereas the green dashed line is a fitted exponential function. The goodness-of-fit test \cite{clauset2009power} gives p-value as $0.58$, which is greater than $0.1$, indicating that the power law is a plausible hypothesis for the data. Vuong's test \cite{vuong1989likelihood} also concludes that the power-law distribution is the closer fitting as the likelihood ratio $LR$ is $5.57$ and the p-value is $2.58\times10^{-8}$, which is less than $0.1$. The AUM data of Willis Towers Watson in 2011, 2015 and 2016 \cite{watson2012world, watson2016world, watson2017world} displays similar results although total AUM expands overall as time goes by. 

\begin{figure}[tbhp]
\begin{center}
\includegraphics[width=.6\textwidth]{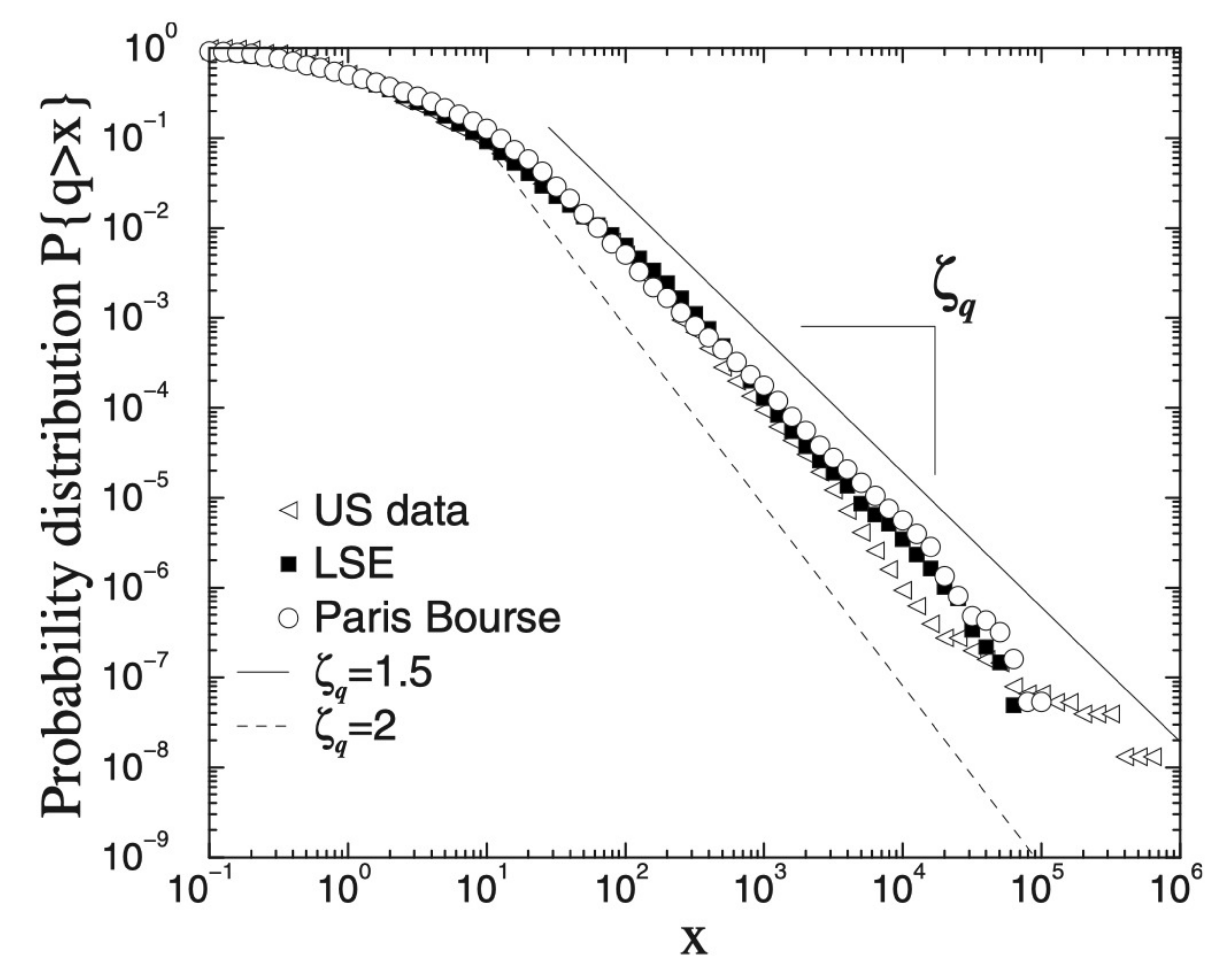}
\end{center}
\caption{Complementary cumulative distribution of the trade size for NYSE (US data), LSE, and Paris Bourse cited from the analysis of Plerou and Stanley \cite{plerou2009reply}. The data for each stock have been scaled by its median. Power-law regressions give $\zeta_q = 1.55 \pm 0.04$ (NYSE), $1.73 \pm 0.02$ (LSE), and $1.60 \pm 0.03$ (Paris Bourse).}
\label{fig6}
\end{figure}

In addition, the emerging mechanism of the volatility clustering found in Speculation Game is also supported by the study of Plerou and Stanley regarding trade size \cite{plerou2009reply}. The authors analyzed tick-by-tick data for three distinct markets: the New York Stock Exchange (NYSE), the London Stock Exchange (LSE), and the Paris Bourse, and reported that the complementary cumulative distribution of individual trading size $q$ displays a power-law tail with the tail exponent $\zeta_q$ as shown in Fig. \ref{fig6}. The slow decay of trade size in this figure implies the existence of a certain number of cash-laden investors. Since $\zeta_q$ is close to $\alpha$ shown in Fig. \ref{fig5}, the power-law nature of trade size seems to be originated from that of AUM. 

\subsection{The role of round-trip trading}
\subsubsection{The horizon of round-trip trades}
\begin{figure}[tbhp]
\begin{center}
\includegraphics[width=.6\textwidth]{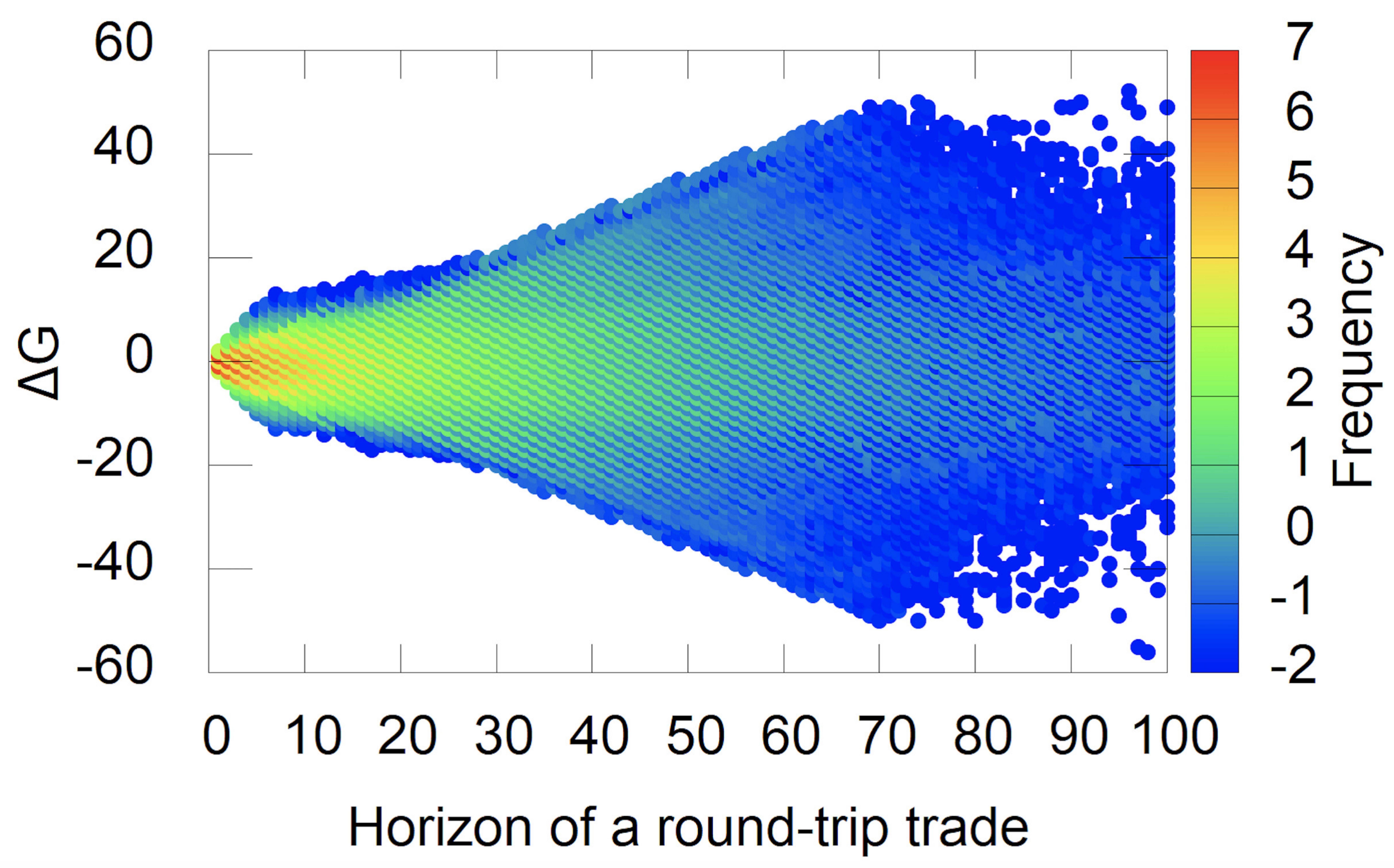}
\end{center}
\caption{The relationship between the strategy gain and the horizon of a round-trip trade along the averaged frequencies in log scale in the case of $M=5$. The color-bar numbers indicate the logarithmic  exponents. The plot is drawn based on the results of round-trip trades completed by all the players from the 100-trials of simulations.}
\label{fig7}
\end{figure}

The reason why such wealthy players can emerge in Speculation Game is related to the trading horizon (or holding period) in the round-trip trades. Fig. \ref{fig7} displays the averaged occurrences of strategy gain resulted from round-trip trades accomplished by all the participated players in the case of $M=5$. Dot color denote those frequencies in the log scale. As shown in Fig. \ref{fig7}, round-trip trades with longer horizons can generate more diverse values of strategy gain, which means that a player tends to win or lose bigger as the horizon of a round-trip trade gets longer. The important point here is the fact that those big losers will either leave (more precisely, be withdrawn from) the market with a high probability or only have scarce wealth even if survived. Thus, the tail of wealth distribution in Fig. \ref{fig4} is built by the wealth of those bigger winners in the round-trip trades.

\begin{figure}[tbhp]
\begin{center}
\includegraphics[width=.6\textwidth]{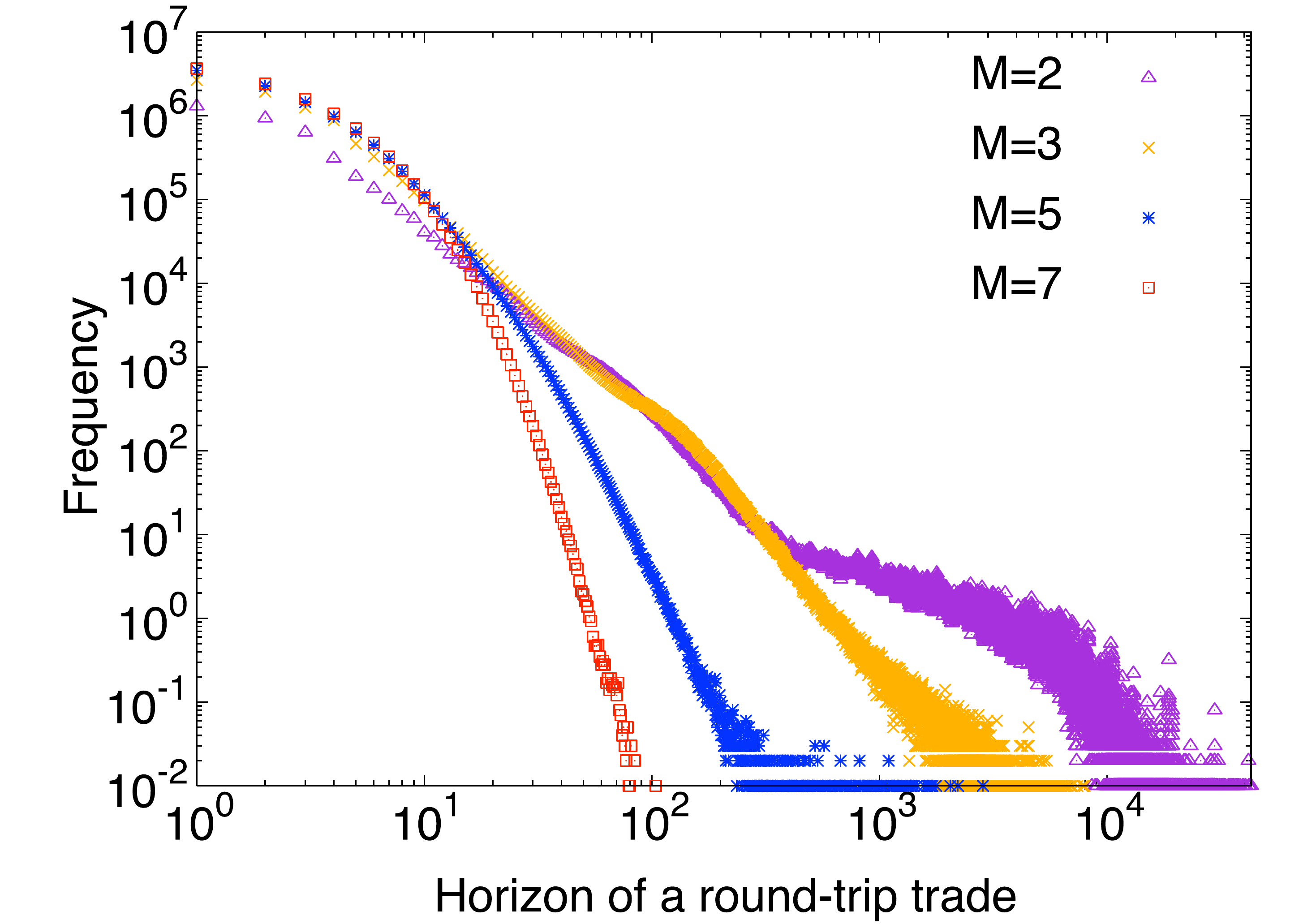}
\end{center}
\caption{The distributions of averaged round-trip trade horizons. These distributions are obtained from round-trip trades completed by all the players from the 100-trials of simulations.}
\label{fig8}
\end{figure}

The log-log plot of averaged distributions of trading horizon are shown in Fig. \ref{fig8}, which displays that the memory size of players affects the horizons of round-trip trades they make. As the figure shows, the blue colored dots have a more spread tail than the red ones have, meaning a greater number of longer trades are made in the case of $M=5$ than those made in the case of $M=7$. Hence, the smaller $M$ the players have, the longer the horizon of a round-trip trade lasts. The situations of $M=2$ and $M=3$ denote the same tendency. The wideness of horizon distributions in Fig. \ref{fig8} corresponds well to that of wealth distributions in Fig. \ref{fig4}. Accordingly, one can conclude that longer round-trip trades would cause the widening of the wealth distribution because of their higher-risk and higher-return nature.

\subsubsection{Empirical evidence}
\begin{figure}[tbhp]
\begin{center}
\includegraphics[width=.6\textwidth]{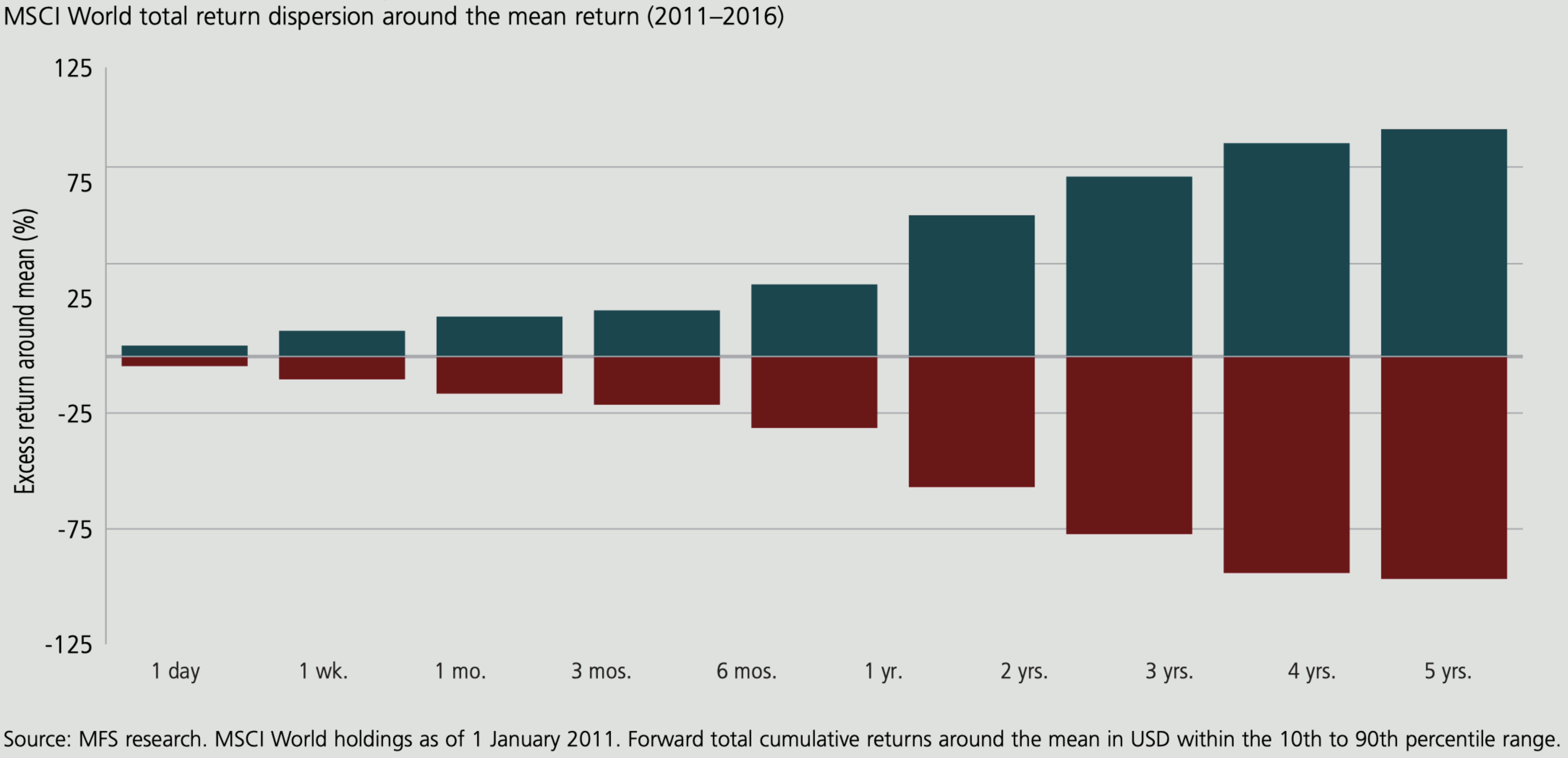}
\end{center}
\caption{The growth of return dispersion with time cited from the analysis of MFS \cite{roberge2017}.}
\label{fig9}
\end{figure}

The diversified values of strategy gain with longer round-trip trades shown in Fig. \ref{fig7} is in accordance with the empirical analysis done by Massachusetts Financial Services (MFS) \cite{roberge2017}. As shown in Fig. \ref{fig9}, MFS reports that the greater return dispersion can be observed as the investment horizon extends, meaning that there are more opportunities for differentiated performance when one holds securities for longer time periods.

\subsubsection{Hold actions}
\begin{figure}[tbhp]
\begin{center}
\includegraphics[width=.6\textwidth]{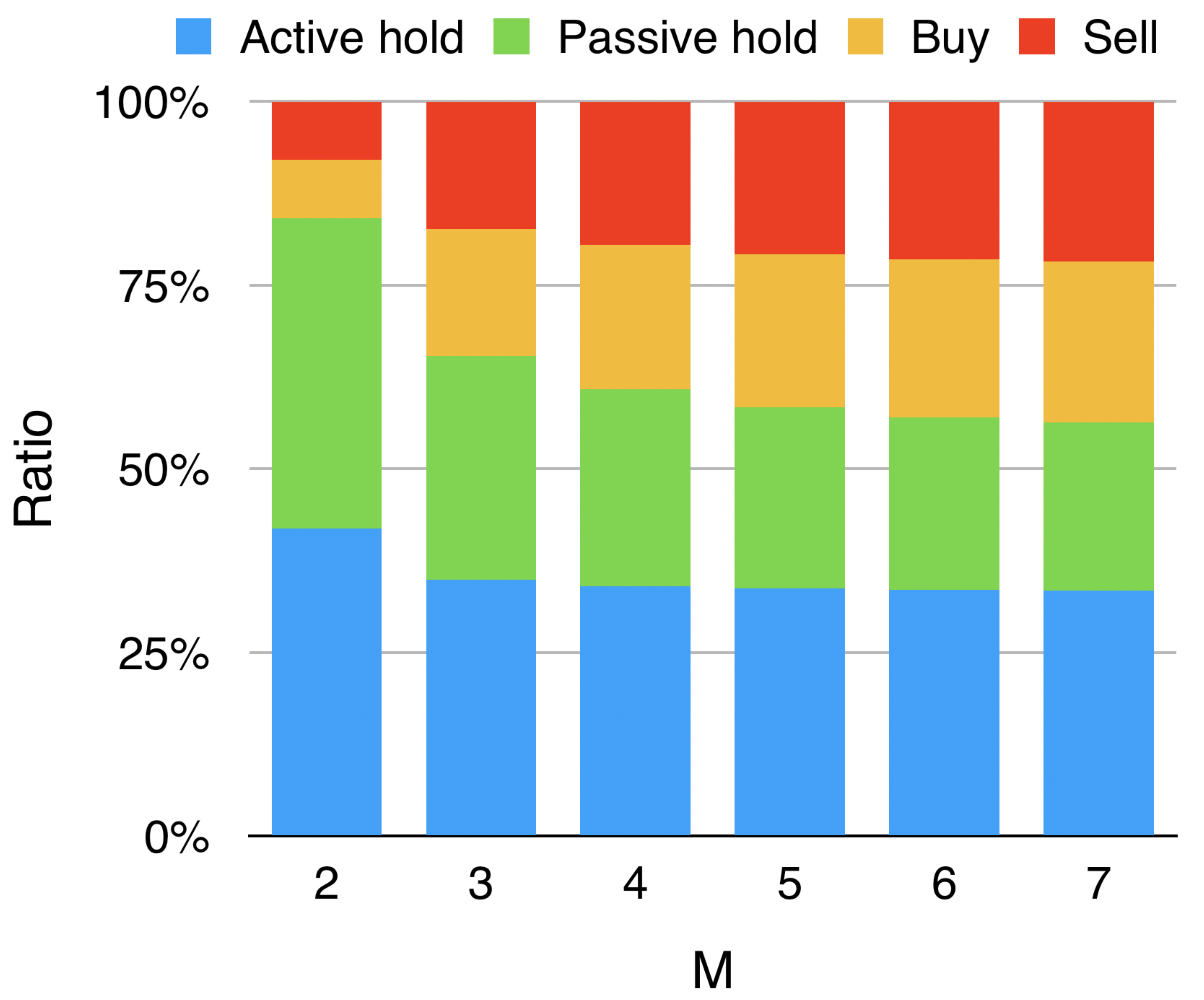}
\end{center}
\caption{The averaged 100\% stacked column chart of four-type actions. This bar graph is drawn from all actions taken by all players from 100-trial simulations.}
\label{fig10}
\end{figure}

The cause of longer trading horizon with smaller $M$ lies with more hold actions taken by the players. In Speculation Game, there are two types of hold action. One type is the {\it active} hold, which is the action a player directly chooses as what her strategy recommends. The other type is the {\it passive} hold, which is the action a player switches to from a prohibited buy or sell action. The prohibition of trade activates when the action recommended by the trading strategy happens to be the same as the one in the opening of position, in order to ensure the completion of a round-trip trade. Fig. \ref{fig10} is the averaged accumulated stick bar graph denoting the statistics of actions among such two types of hold, buy, and sell for all the players along with different memory sizes. As Fig. \ref{fig10} shows, both the ratios of active and passive holds increase as the memory size of players decreases, which correlates to the increment in horizon of a round-trip. Precisely speaking, the ratio of passive hold grows a little larger, indicating the increase of passive hold has a greater effect on making the trading horizon longer. 

It is worth mentioning that, even though the recommendations in strategy tables are generated in a pure random manner, ratios of the four kinds of actions are not balanced especially in the smaller $M$ cases, which indicates the loss of the ergodicity in the system. If the system is ergodic, the summed ratio of passive hold, buy and sell should be double to the ratio of active hold , as the lager $M$ cases shown in Fig. \ref{fig10}. 

\subsection{Other factors}
\subsubsection{Signal dependency}
\begin{figure}[tbhp]
\begin{center}
(a)\includegraphics[width=0.85\textwidth]{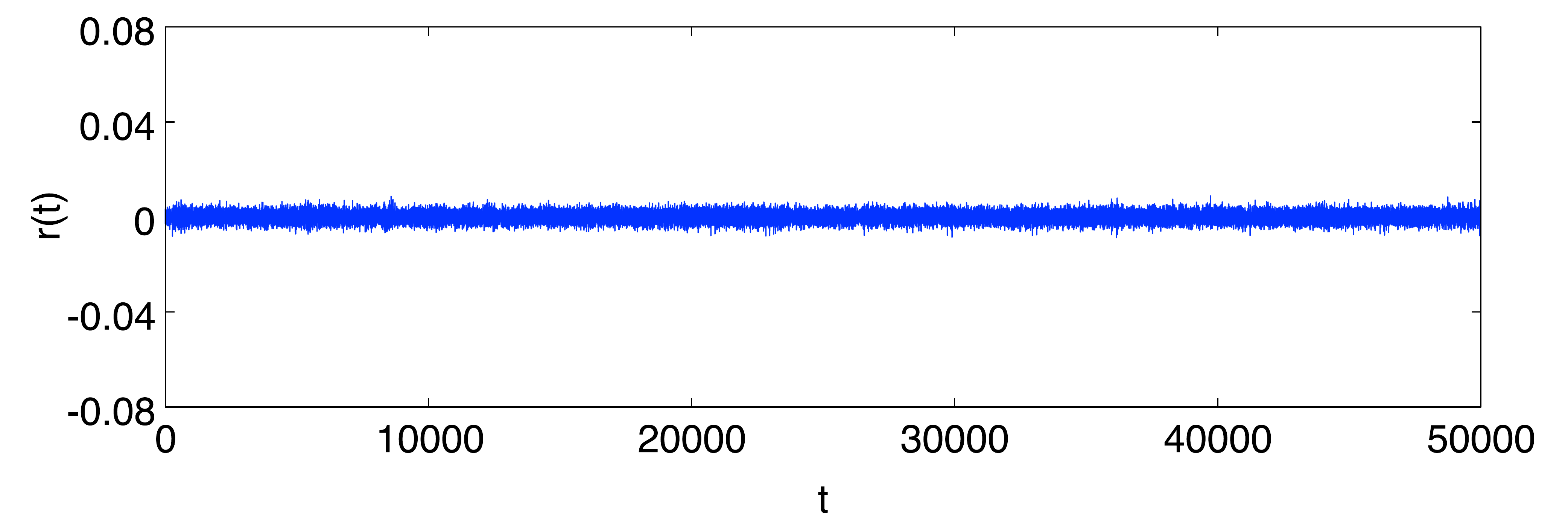}
(b)\includegraphics[width=0.85\textwidth]{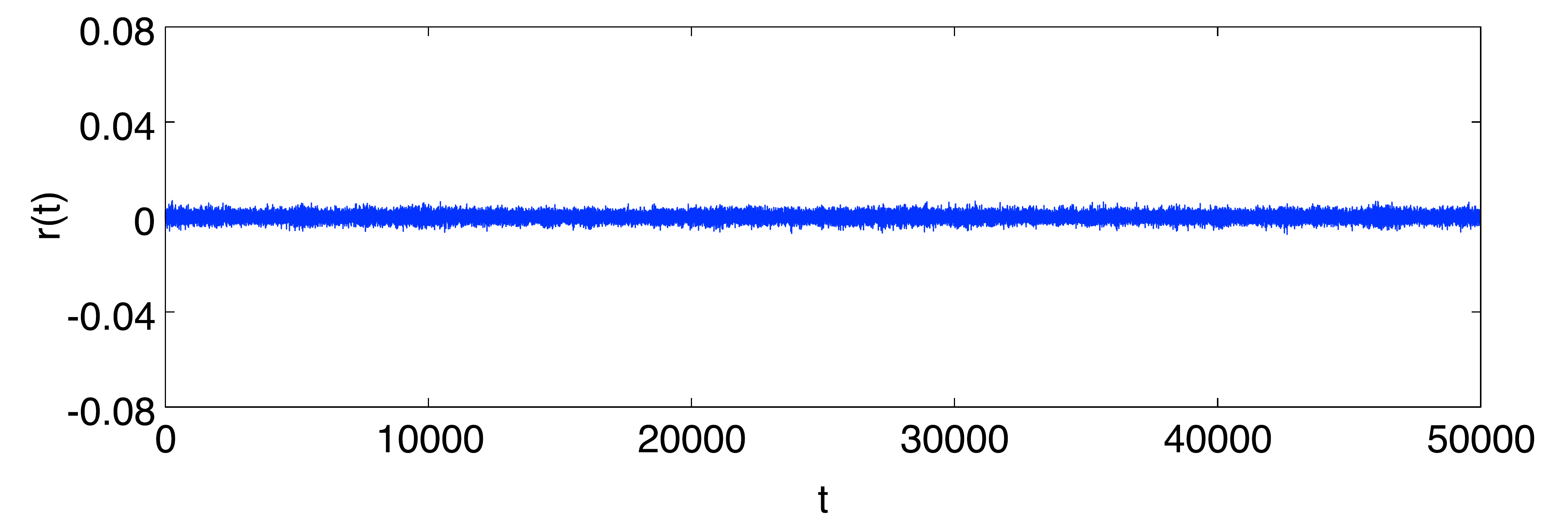}
\end{center}
\caption{Time series of market returns in the cases (a) when an external common history is randomly given and (b) when a buy or sell position is randomly opened with $p=0.5$ ($N=1,000$, $M=5$, $S=2$, $B=9$, $C=3$).}
\label{fig11}
\end{figure}

The non-uniformity in the ratio of trading actions is related to the fact that Speculation Game is a highly signal dependent game. If the genuine historical pattern were not employed, the game would not work properly. In contrast, an exogenous random history works well in Minority Game as long as it is common to all the participants of the game \cite{cavagna1999irrelevance}. In Fig. \ref{fig11}(a), a time series generated by Speculation Game while employing the exogenous histories is shown. It is obvious that the data does not contain any clustered temporal structures, which also supports findings from the Patzelt-Pawelzik model which states that, only the endogenously generated news (history), rather than the external one,  can account for the large price fluctuations \cite{patzelt2013inherent}. Another demonstration of the failure in the generation of volatility clustering is the case in which a round-trip trade was randomly opened with a probability $p=0.5$ without referencing the price history as well as the current position. As panel (b) of Fig. \ref{fig11} displays, similarly to the result in panel (a), there is no evidence for the intermittent burst of price fluctuations. From these demonstrations, one can conclude that the decision-making process of players for the round-trip trading, which results in the nonergodic actions, is indispensable to generate the volatility clusteringin Speculation Game. In fact, round-trip trading can also generate distinctive historical patterns of price returns in Speculation Game, which will be detailed in our future report.

\subsubsection{Inductive learing}\label{learning}
\begin{figure}[tbhp]
\begin{center}
(a)\includegraphics[width=.6\textwidth]{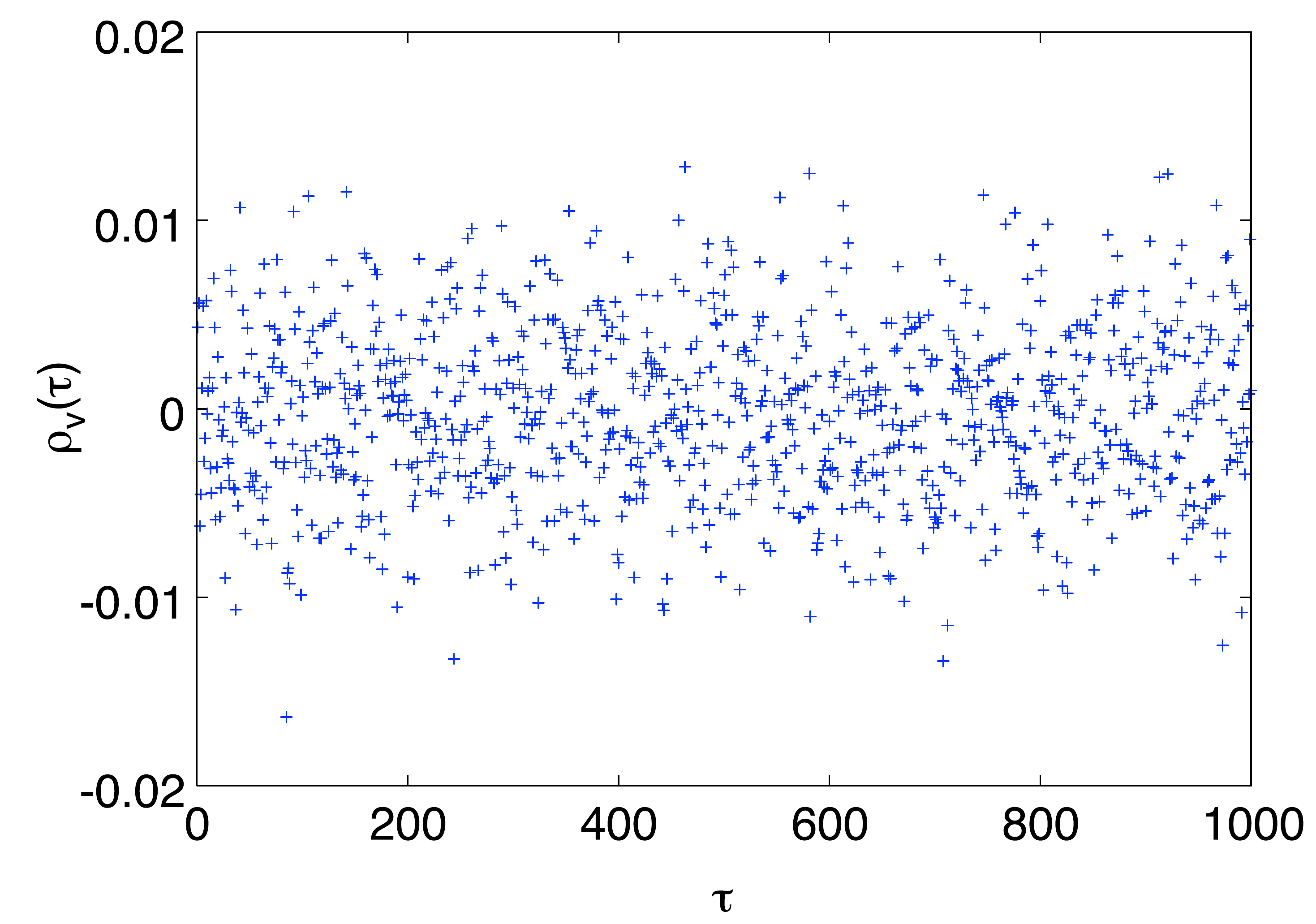}
(b)\includegraphics[width=.6\textwidth]{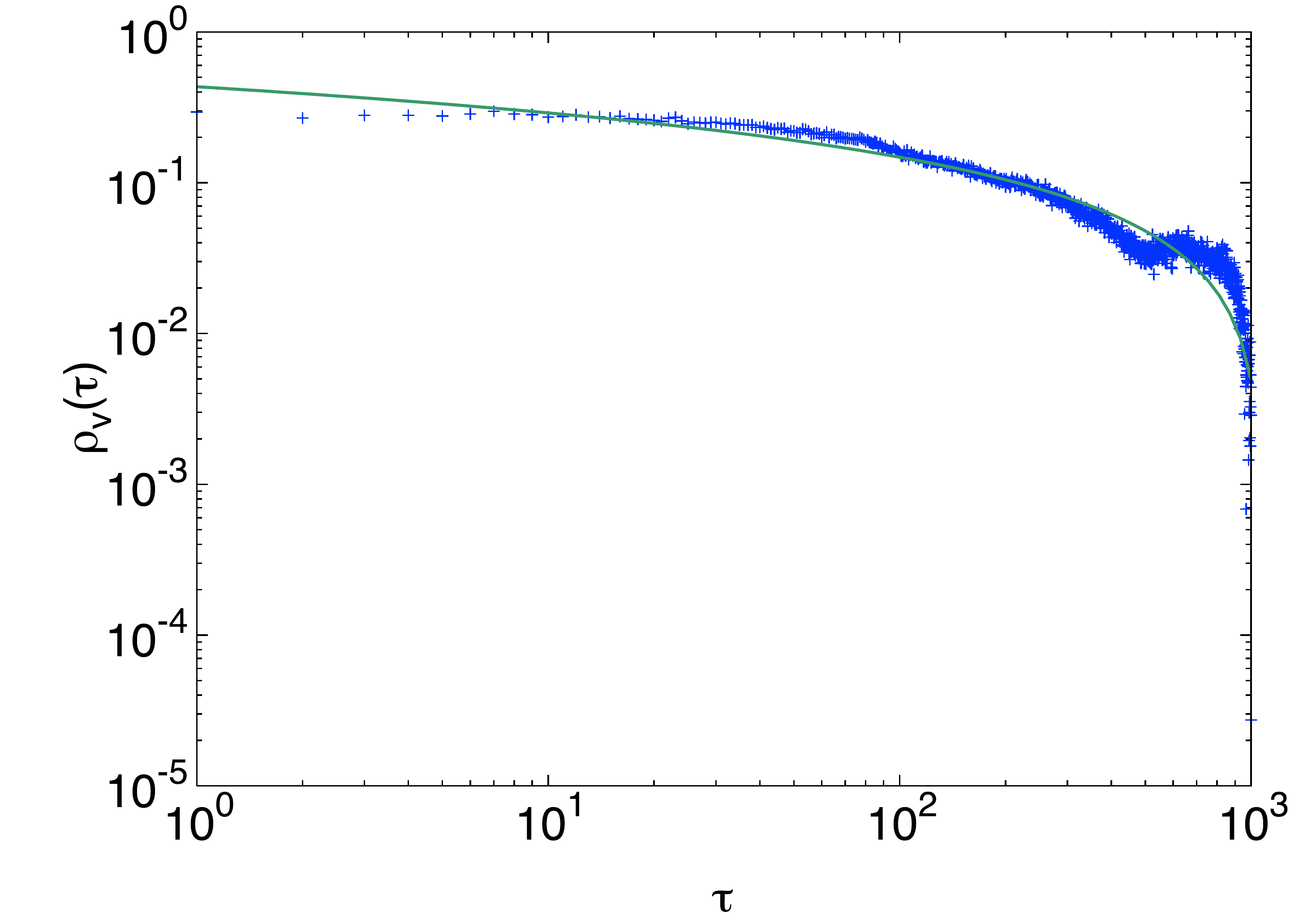}
\end{center}
\caption{(a) The absence of autocorrelation in volatility in MGDC in the case of $S=1$. The rest parameters are set as $P=260$, $N_s=401$, $N_p=401$, $\epsilon=0.01$ by referring \cite{challet2001minority} and \cite{galla2009minority}. (b) The slow decay of autocorrelation in volatility in Speculation Game in the case of $S=1$ ($N=1,000$, $M=5$, $B=9$, $C=3$). The decay can be regressed to a logarithmic function: $\rho_v(\tau)=-0.062{\rm ln}(\tau)+0.433$ ($R^2=0.951$), as the green line shows.}
\label{fig12}
\end{figure}

Lastly, it should be remarked that the structure of round-trip trading could be a sole crucial factor for the emergence of volatility clustering. In Minority Game typed models, the inductive learning is needed for the elicitation of herding behavior among agents. If agents owned only one strategy, the function of inductive learning would not work in these models. As a typical example, in MGDC\cite{galla2009minority} (an extended Minority Game model equipped with dynamic capital and variable investments) with a single strategy table for each agent, the slow decay of autocorrelation in volatility measured by
\begin{equation}
\rho_v(\tau) = {\rm Corr}(|r(t+\tau)|, |r(t)|),
\label{eq9}
\end{equation}
where $|r(t)|$ is volatility and $\tau$ is the time lag, will be absent due to the loss of adaptivity derived from the inductive learning. As panel (a) of Fig. \ref{fig12} shown, the autocorrelation in volatility drops immediately to the noise level in MGDC with a single trading strategy ($S=1$) for each player. To recover its reproducibility, MGDC agents typically need more than one trading strategy to grant their adaptability for the dynamical market circumstances \cite{challet2001minority}. 

On the contrary, inductive learning is not indispensable for the recovery of long-range autocorrelation of volatility in Speculation Game. Even in the case when players behave mechanically, namely $S=1$, the autocorrelation of volatility decays slowly similar to the cases of $S\geq2$, as the log-log plot in panel (b) of Fig. \ref{fig12} displays \footnote{See Appendix C for the effect of $S$ as well as $C$ on $\Delta p$.}. The remarkable gap between Fig. \ref{fig12}(a) and Fig. \ref{fig12}(b) elucidates the importance of round-trip trading itself in Speculation Game for the emergence of long-range autocorrelation of volatility. More explicitly speaking, while a combination of defined round-trip trading and dynamic wealth is enough for the reproduction of those stylized facts in Speculation Game, an employment of inductive learning and evolving capital is essentially needed in those Minority Game type models. 

\section{Conclusion}
The analysis of Speculation Game with the simulations and the data from the actual financial markets illustrates that the emergence of volatility clustering can be induced by the relatively big and medium order placements from the activated rich players, when the market environment is suitable for an appropriate number of such players. The spontaneous redistribution of market wealth through repetitions of round-trip trades widens the wealth disparity among the players and form a heterogeneous distribution, which can be regarded as another possible origin (other than the herding behavior) of the volatility clustering in the price return. This result is in accordance with the thinking elucidated in the previous study \cite{takayasu1990fractal}, which states that fractal structures in stock price change are caused actually by that rich and poor traders make decisions essentially in the same way even though they place very different amounts of trading unit. Furthermore, factors influencing the wealth disparity, such as the relationship between the strategy gain and the trading horizon in a round-trip trade, the significance of active and passive holds for the nonergodic decision making, and the strong dependency of the system on the genuine historical signals, are clarified. Finally, the comparison with MGDC reveals that the structure of round-trip trading is the critical element for Speculation Game to reproduce the long-range autocorrelation of volatility.

\section*{Acknowledgments}
This work was supported by JSPS KAKENHI grant number JP17J09156.

\appendix
\setcounter{figure}{0}
\section{Time series in the cases of $M=2$ and $M=3$}
\begin{figure}[tbhp]
\begin{center}
(a)\includegraphics[width=0.85\textwidth]{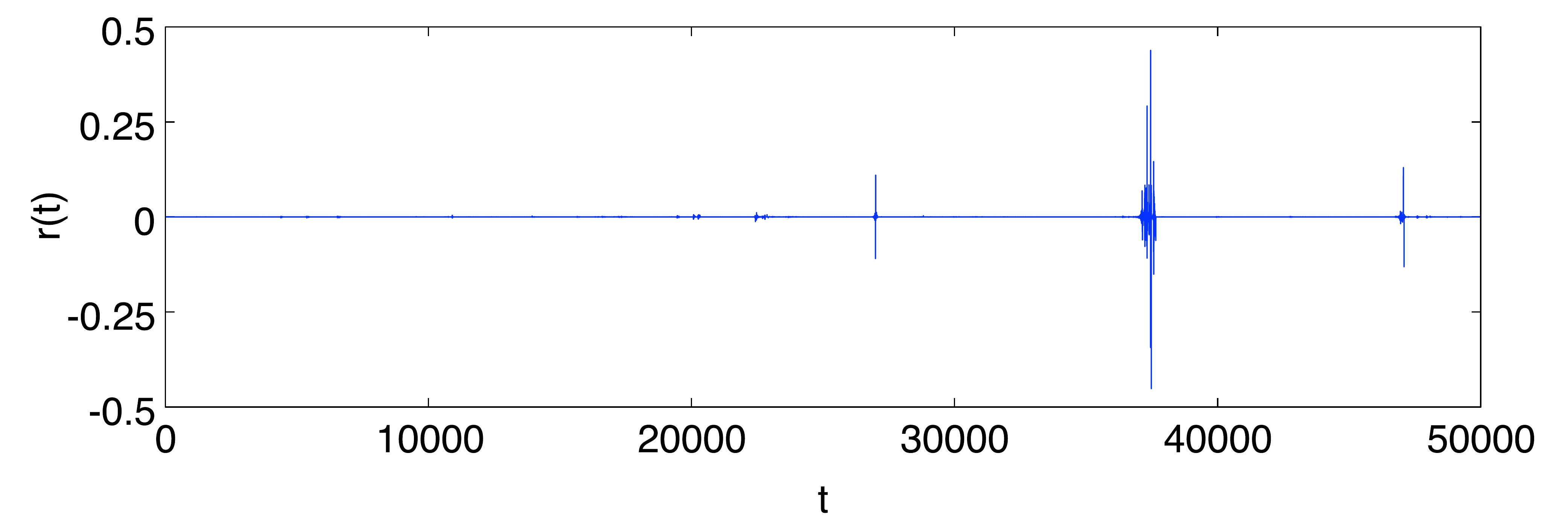}
(b)\includegraphics[width=0.85\textwidth]{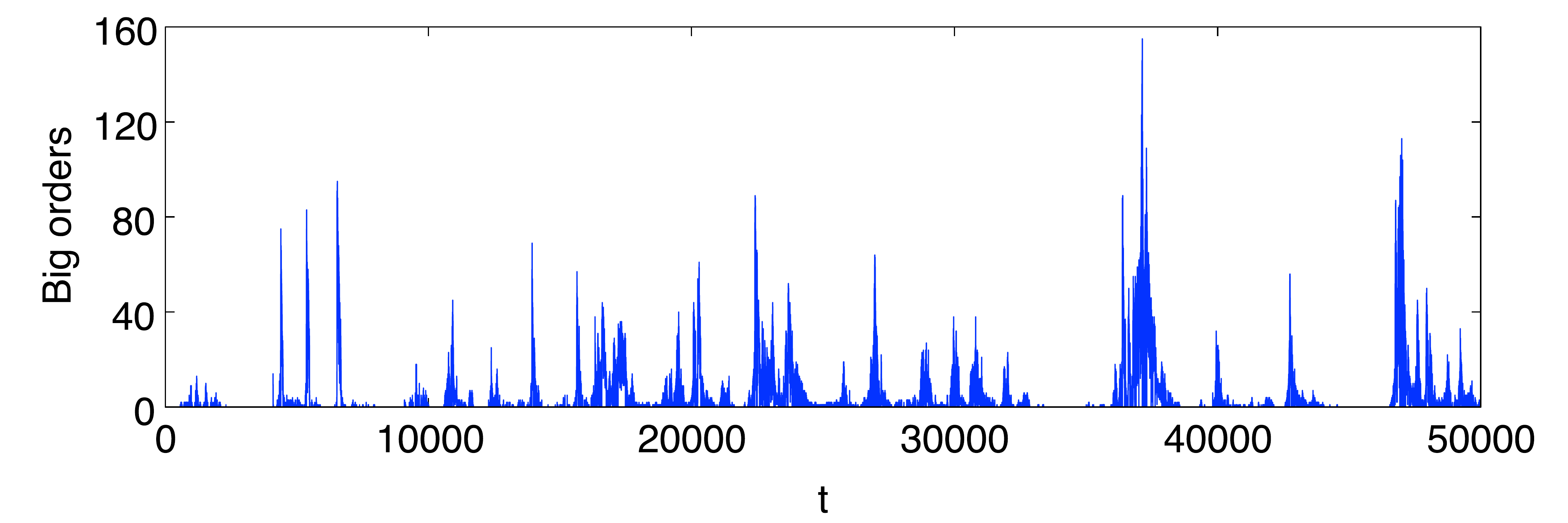}
(c)\includegraphics[width=0.85\textwidth]{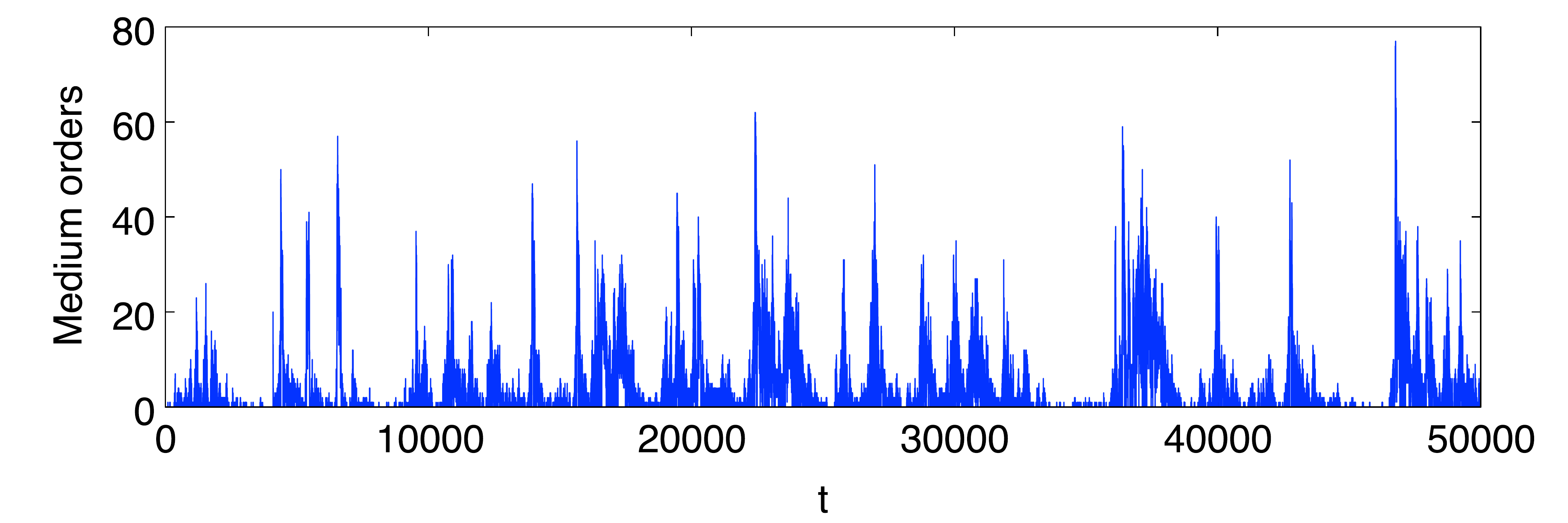}
(d)\includegraphics[width=0.85\textwidth]{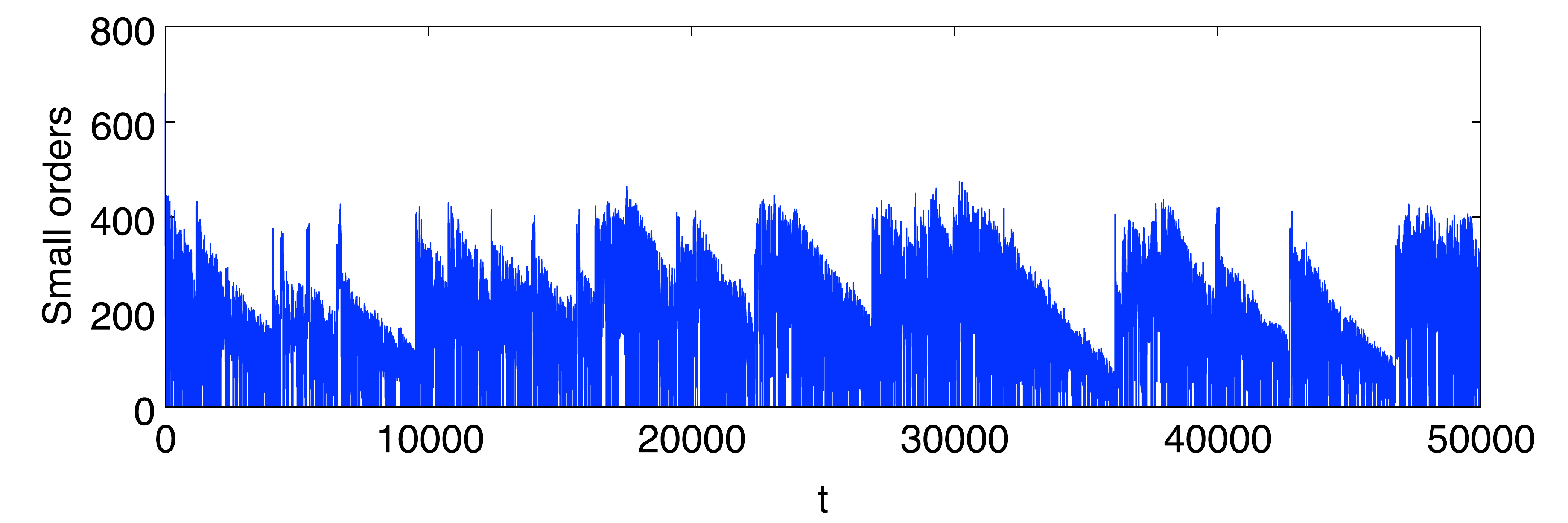}
\end{center}
\caption{Time series of (a) market returns and of the number of (b) big orders, (c) medium orders, and (d) small orders in the case of $M=2$ ($N=1,000$, $S=2$, $B=9$, $C=3$). The initial market price is set as $p(0)=10,000$ to calculate $r(t)$.}
\label{fig13}
\end{figure}

In the case of $M=2$, the system frequently reaches the extreme state involving tremendous bursts of market return $r(t)$ as the panel (a) of Fig. \ref{fig13} displays. These bursts are brought by a great number of extremely large order volumes. As the purple dots ($M=2$) show in Fig. \ref{fig4}, the wealth of some players can even reach $10^5$. Those super wealthy players place enormous quantities because there is no limitation for trading volumes. Moreover, there are overall much more big (and medium) orders throughout the simulation, as shown in the panel (b) (and (c)) of Fig. \ref{fig13}. Lastly, the most distinctive characteristic of the extreme state is the alternative appearances of constant decrease and rapid recovery in small orders, as shown in the panel (d) of Fig. \ref{fig13}.

\begin{figure}[tbhp]
\begin{center}
(a)\includegraphics[width=0.85\textwidth]{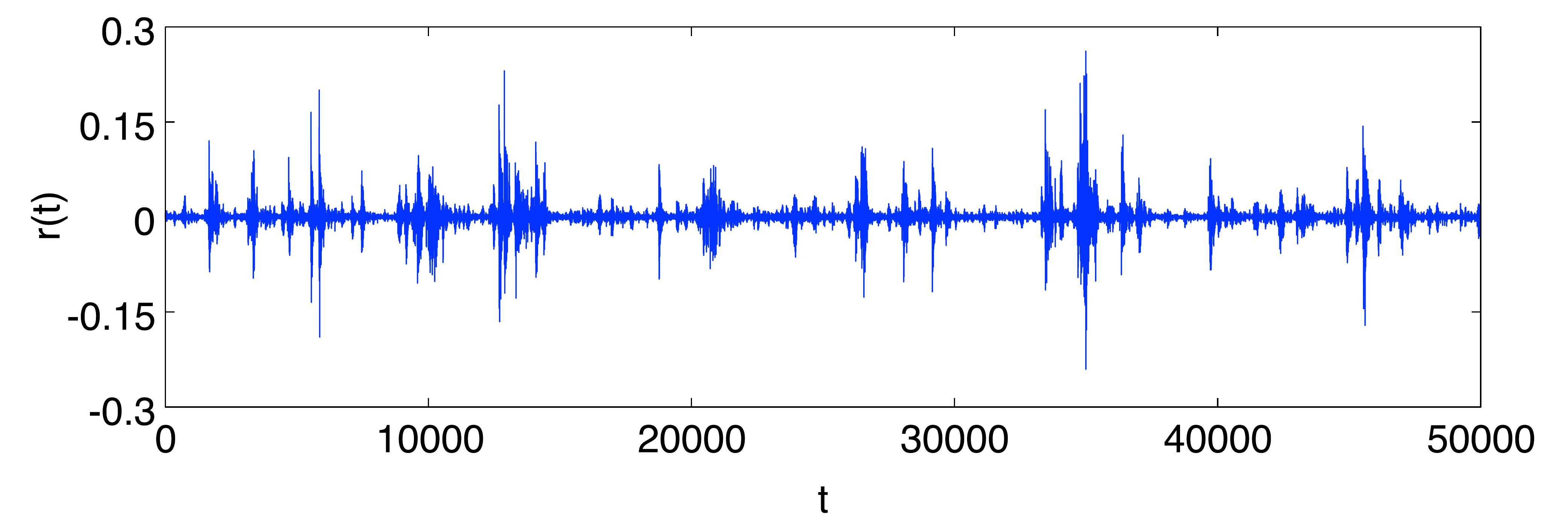}
(b)\includegraphics[width=0.85\textwidth]{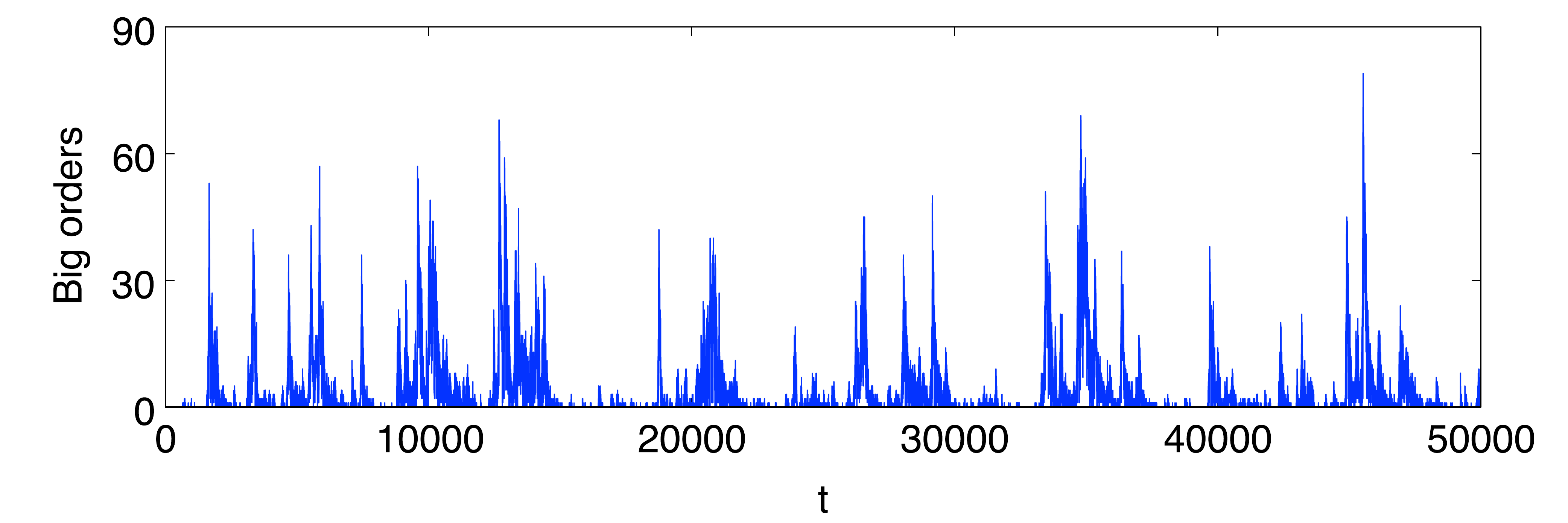}
(c)\includegraphics[width=0.85\textwidth]{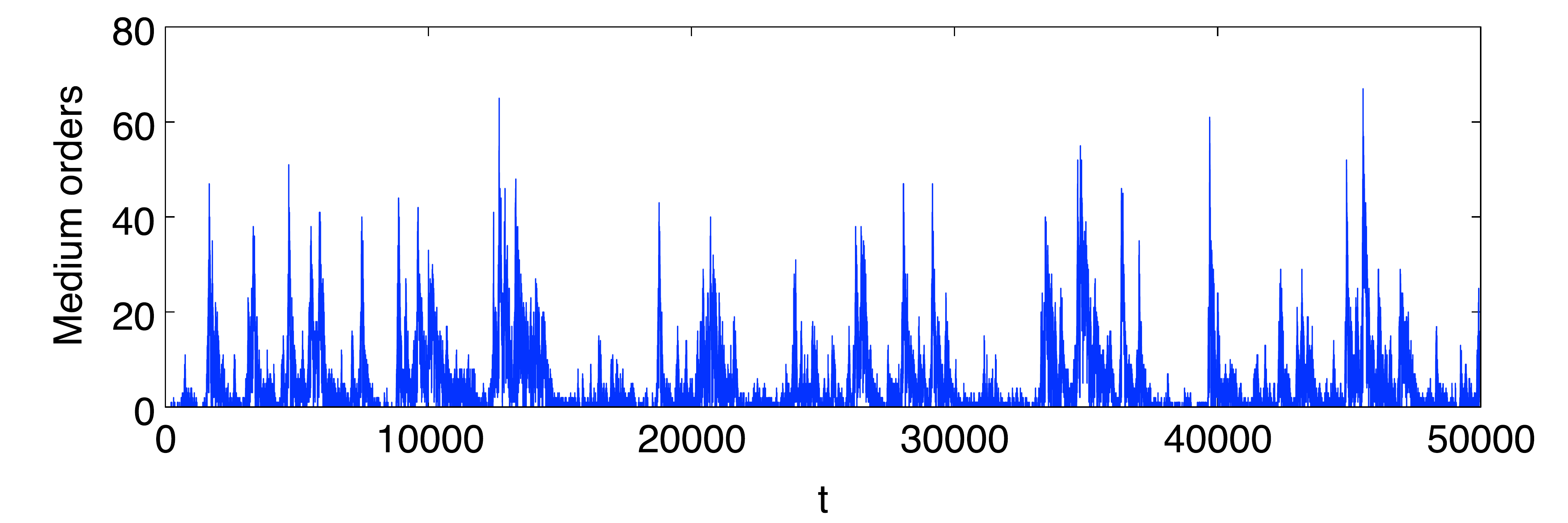}
(d)\includegraphics[width=0.85\textwidth]{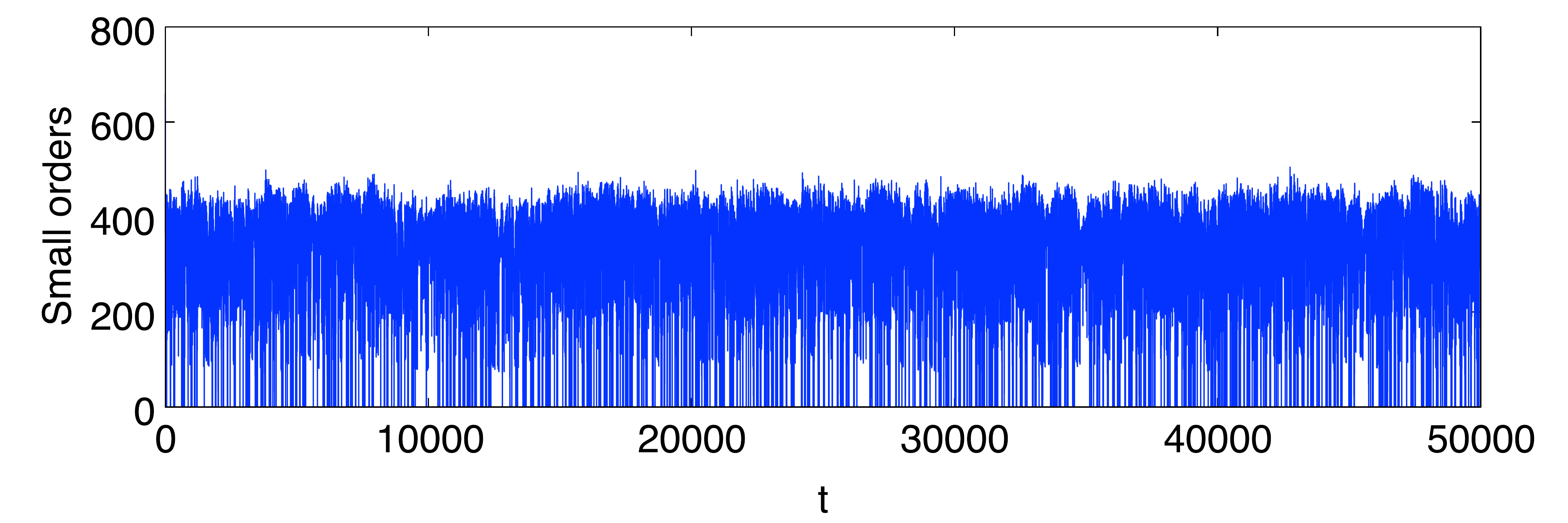}
\end{center}
\caption{Time series of (a) market returns and of the number of (b) big orders, (c) medium orders, and (d) small orders in the case of $M=3$ ($N=1,000$, $S=2$, $B=9$, $C=3$).}
\label{fig14}
\end{figure}

In the case of $M=3$, the market return $r(t)$ exhibits much stronger clustering of volatilities in comparison with the case of $M=5$ as the panel (a) of Fig. \ref{fig14} displays. This is also due to the existence of larger number of wealthier players (see the orange dots in Fig. \ref{fig4}), which could result in more placements of big and medium orders through the whole time periods, as shown in the panel (b) and (c) of Fig. \ref{fig14}. The incessant presence of small orders can be observed but with a bit more fluctuations and more vacant points (the panel (d) of Fig. \ref{fig14}) than in the cases of $M=5$ and $M=7$.

\section{The wider wealth disparity with smaller $B$}
\begin{figure}[tbhp]
\begin{center}
\includegraphics[width=.6\textwidth]{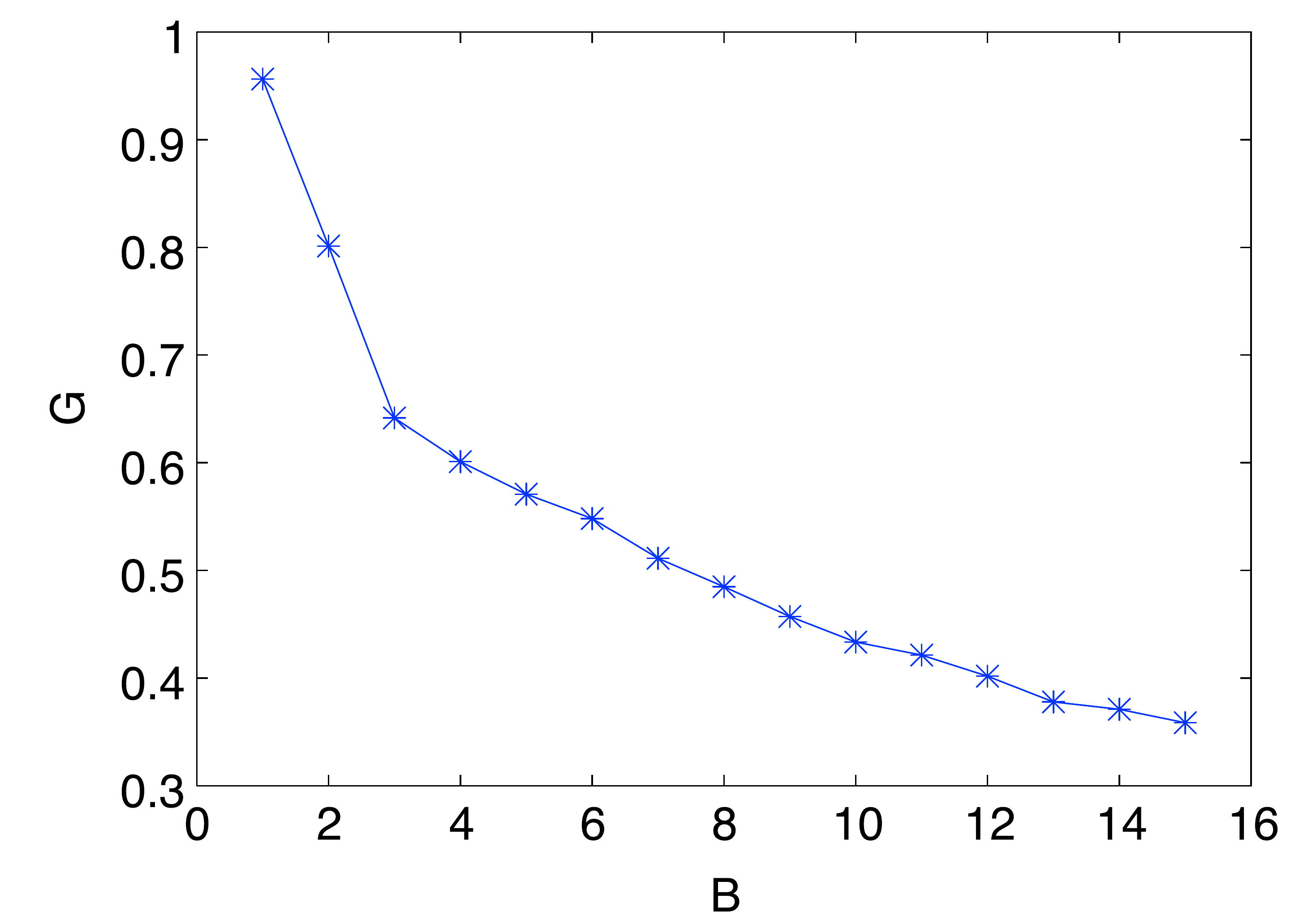}
\end{center}
\caption{The relationship between $B$ and the 100-trial averaged Gini coefficient at $t=50,000$ ($N=1,000$, $M=5$, $S=2$, $C=3$).}
\label{fig15}
\end{figure}

Smaller $B$, meaning more liquidity with the easier placement of multiple quantities, can also widen the wealth disparity between the rich and the poor. The wealth inequality among players can be measured by the Gini coefficient, which is defined as follows,
\begin{equation}
G = 1 - 2\int_{0}^{1}L(u)du,
\label{eq10}
\end{equation}
where $L(u)$ is the Lorenz curve \cite{dorfman1979formula}. Fig. \ref{fig15} is a snapshot of 100-trial averaged Gini coefficient calculated at $t=50,000$ with various values of $B$. In Fig. \ref{fig15}, $G$ increases as $B$ decreases, meaning smaller $B$ widens the wealth differential among players, which corresponds to the increase of $\sigma$ with the reduction of $B$ in Fig. \ref{fig1}. Moreover, in the cases of $B<3$, the growth rate of $G$ increases dramatically, corresponding with the market price change in the extreme states. Namely, without any limitation for the trading volumes, the excessive liquidity can collapse the system as the extreme state reached.

\section{The roiling effect of $S$ and $C$}
\begin{figure}[tbhp]
\begin{center}
\includegraphics[width=.6\textwidth]{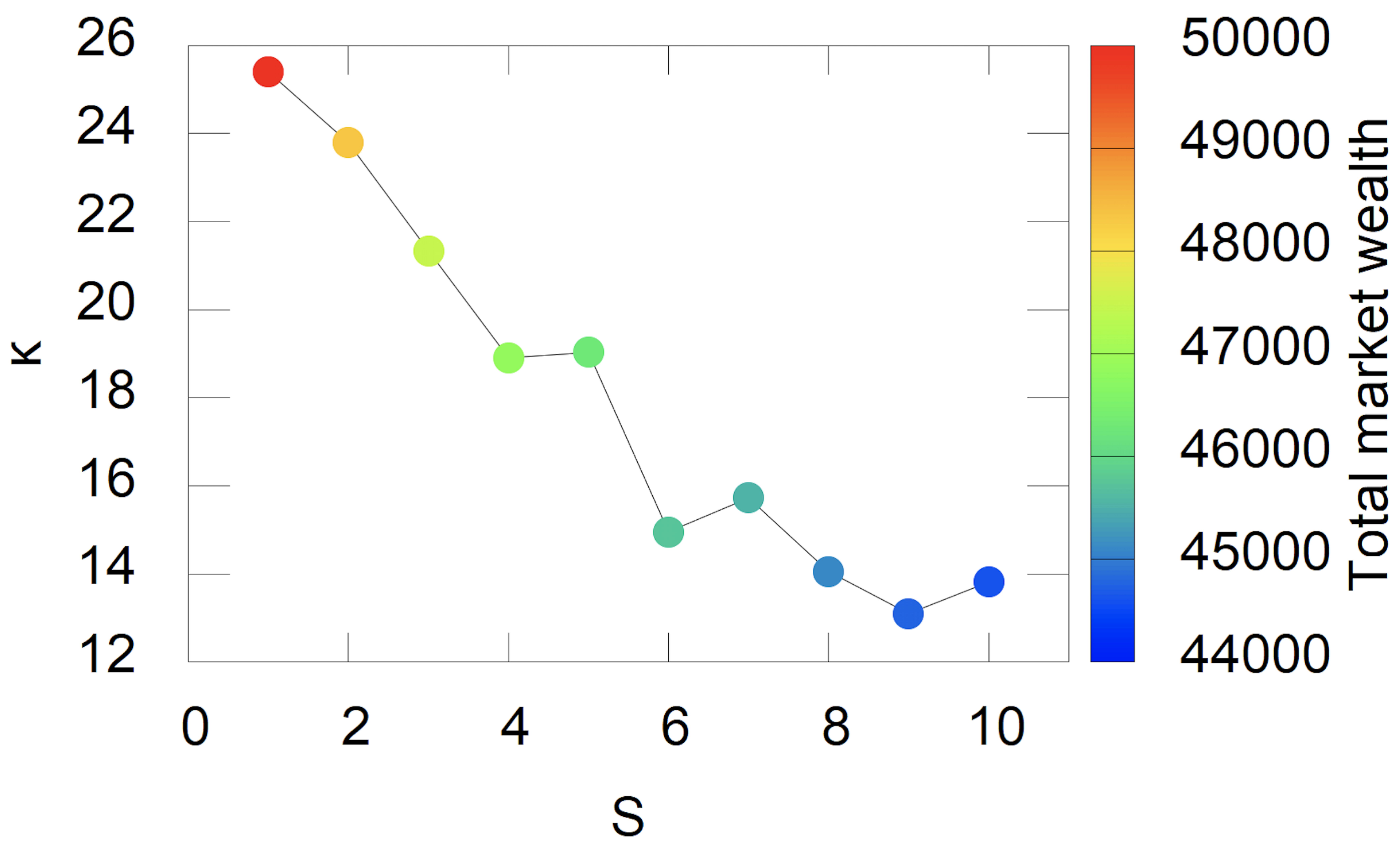}
\end{center}
\caption{The decay of averaged kurtosis of $\Delta p$ as $S$ increases ($N=1,000$, $M=5$, $B=9$, $C=3$). The dot color shows the time-averaged total market wealth of all players. The plot is based on the 100-trials of simulations.}
\label{fig16}
\end{figure}

A larger number of strategy would weaken the intensity of clustered volatility and makes the tails of $\Delta p$ distribution less heavy. The heaviness of tails can be measured by the kurtosis of market price change, which is defined as
\begin{equation}
\kappa = \frac{\langle (\Delta p(t) - \langle \Delta p(t) \rangle)^4 \rangle}{\sigma^4}-3.
\label{eq11}
\end{equation}
Fig. \ref{fig16} shows the relationship between $S$ and the 100-trial averaged $\kappa$, indicating $\kappa$ decreases as $S$ increases. This is because that larger $S$ could raise the possibility of switching the strategy in use, which would further induce a stronger scrambling effect on market conditions. In other words, a greater number of strategy switching can accelerate the change of market states and make players more difficult to gain profits through round-trip trades, thus render the emergence of big players harder. Meanwhile, the degree of market roiling can be measured by the statistics of the time-averaged total market wealth of all players through the whole period, as the dot color in Fig. \ref{fig16} displays. The gradual transition of dot colors from red to blue indicates that the time-averaged total market wealth decreases as $S$ becomes bigger, which validates the increase of difficulty to make money through round-trip trades. Note this result resembles the results obtained in Minority Game which shows that a greater number of strategies (or a higher switching rate of strategies) lowers the success rate \cite{challet1997emergence};

\begin{figure}[tbhp]
\begin{center}
\includegraphics[width=.6\textwidth]{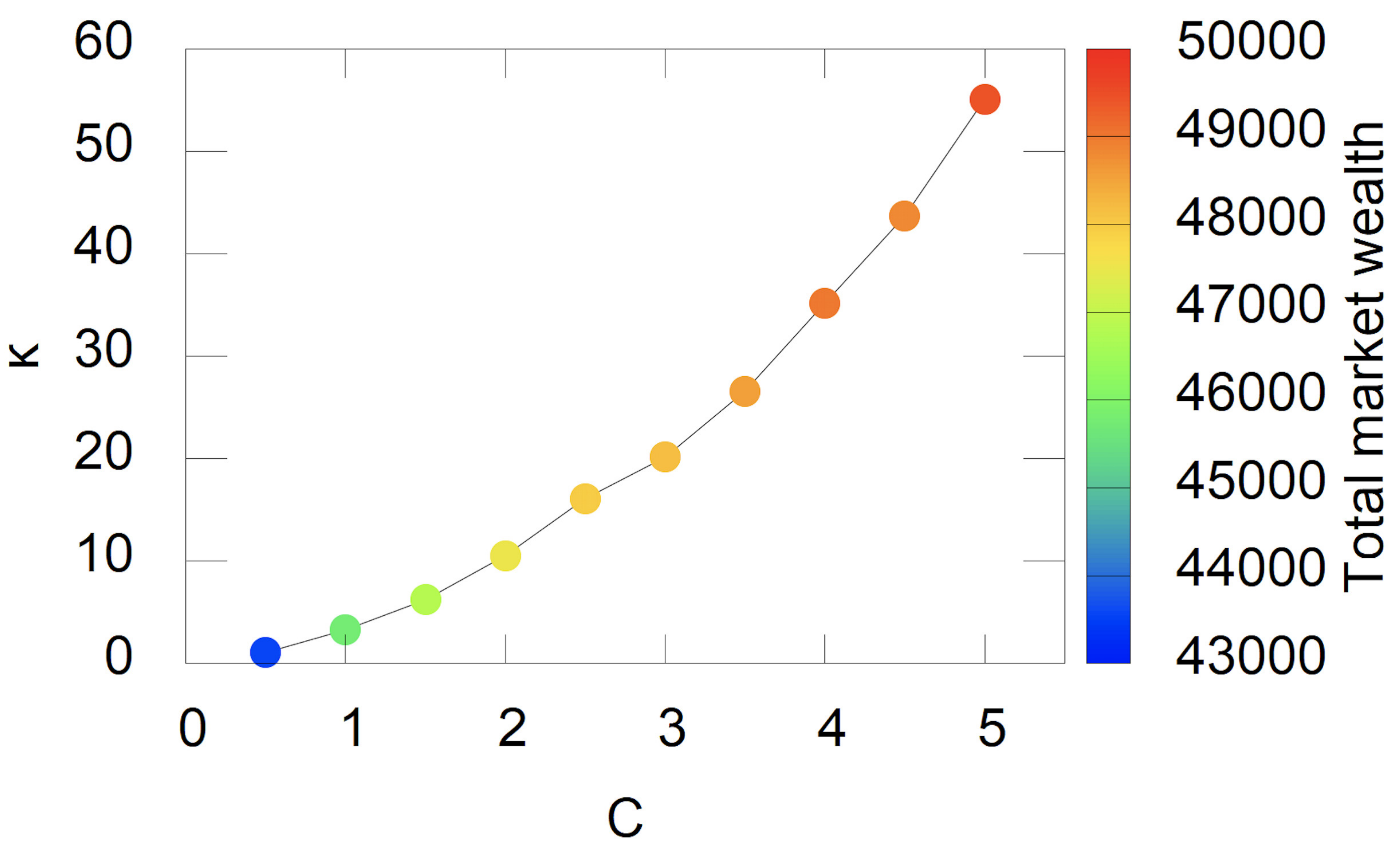}
\end{center}
\caption{The growth of averaged kurtosis of $\Delta p$ as $C$ increases ($N=1,000$, $M=5$, $S=2$, $B=9$). The dot color shows the time-averaged total market wealth of all players. The plot is based on the 100-trials of simulations.}
\label{fig17}
\end{figure}

A smaller cognitive threshold $C$ can also suppress large fluctuations, so that the averaged $\kappa$ decreases as $C$ decreases in Fig. \ref{fig17}. According to the definition of $C$ in Eq. \ref{eq4}, when $C$ is smaller, histories including the signals of $2$ and $-2$ appear more frequently and the corresponding trading recommendations in the strategy tables are activated more frequently. Hence, the cognitive threshold can roil the market like the cases of higher strategy switching rate. Indeed, the time-averaged total market wealth decreases as $C$ decreases, as the gradual change of dot colors in Fig. \ref{fig17} shows. 

\bibliography{mybibfile}

\end{document}